# Establishing correspondence between the reformulation of quantum mechanics without a potential function and the conventional formulation

A. D. Alhaidari

*Saudi Center for Theoretical Physics, P.O. Box 32741, Jeddah 21438, Saudi Arabia*

**Abstract**: Within the recent reformulation of quantum mechanics where a potential function is not required, we show how to reconstruct the potential so that a correspondence with the standard formulation could be established. However, severe restriction is placed by the correspondence on the kinematics of such problems.

We are honored to dedicate this work to Prof. Hashim A. Yamani on the occasion of his 70th birthday.

## 1. Introduction

Recently, we introduced a formulation of quantum mechanics without the need to specify a potential function [1-3]. This was done under the assumption that there are quantum systems that could be described analytically but their potential functions are either difficult to specify or impossible to realize analytically. For example, these potential functions might be nonanalytic, nonlocal, energy dependent, or the corresponding differential wave equation is higher than second order (possibly of infinite order) and so on. The success of this formulation implies that the class of exactly solvable problems is larger than the well-known class in the conventional formulation of quantum mechanics (see, for example, Ref. [4] for the conventional class of potential functions). In the new formulation, the wavefunction is expanded over a complete set of square-integrable basis functions in configuration space. The expansion coefficients are orthogonal polynomials in the energy. Specifically, we write the total wavefunction as $\Psi(t,x)=e^{-iEt}\psi(E,x)$, where $E$ is the energy of the system in the atomic units $\hbar=m=1$ and [1-3]

$$\psi(E,x)=\sqrt{\rho^{\mu}(\varepsilon)}\sum_{n}P_n^{\mu}(\varepsilon)\phi_n(x), \tag{1}$$

which is assumed to be a bounded sum. $\{\phi_n(x)\}$ is the complete set of square-integrable basis functions in configuration space with coordinate $x$ and $P_n^{\mu}(\varepsilon)$ is an orthogonal polynomials of order $n$ in the variable $\varepsilon$ which is some proper function of the energy. The symbol $\mu$ stands for a set of real parameters associated with the physical system and $\rho^{\mu}(\varepsilon)$ is the normalized weight function associated with these energy polynomials. That is, $\int\rho^{\mu}(\varepsilon)P_n^{\mu}(\varepsilon)P_m^{\mu}(\varepsilon)\,d\eta(\varepsilon)=\delta_{nm}$, where $d\eta(\varepsilon)$ is an appropriate measure. The energy polynomials also satisfy a symmetric three-term recursion relation as follows

$$\varepsilon\,P_n^{\mu}(\varepsilon)=a_n^{\mu}P_n^{\mu}(\varepsilon)+b_{n-1}^{\mu}P_{n-1}^{\mu}(\varepsilon)+b_n^{\mu}P_{n+1}^{\mu}(\varepsilon),\qquad n=1,2,... \tag{2}$$

where $P_0^\mu(\varepsilon)=1$, $P_1^\mu(\varepsilon)=\alpha\,\varepsilon+\beta$ and $\left\{a_n^\mu,b_n^\mu\right\}$ are the recursion coefficients with $b_n^\mu\neq 0$ for all $n$. If $\alpha=1/b_0^\mu$ and $\beta=-a_0^\mu/b_0^\mu$ , then we refer to $\left\{P_n^\mu(\varepsilon)\right\}$ as "polynomials of the first kind". The basis set $\left\{\phi_n(x)\right\}$ does not contain any structural or dynamical information about the physical system under study. They satisfy the boundary conditions and contain only kinematical information such as the angular momentum and a length scale. On the other hand, full physical information about the system is contained only in the energy polynomials $\left\{P_n^\mu(\varepsilon)\right\}$ and the corresponding weight function $\rho^\mu(\varepsilon)$. The physically relevant energy polynomials are those with the following asymptotics (limit as $n\to\infty$) [1-3]

$$P_n^\mu(\varepsilon)\approx n^{-\tau}A^\mu(\varepsilon)\times\cos[n^\xi\,\theta(\varepsilon)+\delta^\mu(\varepsilon)], \tag{3}$$

where $\tau$ and $\xi$ are real positive constants that depend on the particular polynomial. $A^\mu(\varepsilon)$ is the scattering amplitude and $\delta^\mu(\varepsilon)$ is the phase shift. Bound states, if they exist, occur at (infinite or finite) energies that make the scattering amplitude vanish. That is, the $m^{\text{th}}$ bound state occurs at an energy $E_m$ such that $A^\mu(\varepsilon_m)=0$ and the corresponding bound state is written as

$$\psi(E_m,x)=\sqrt{\omega^\mu(\varepsilon_m)}\sum\nolimits_n Q_n^\mu(\varepsilon_m)\phi_n(x), \tag{4}$$

where $\left\{Q_n^\mu(\varepsilon_m)\right\}$ are the discrete version of the polynomials $\left\{P_n^\mu(\varepsilon)\right\}$ and $\omega^\mu(\varepsilon_m)$ is the associated discrete weight function. In the absence of a potential function, the physical properties of the system are deduced from the properties of the associated orthogonal polynomials. Such properties include the shape of the weight function, nature of the generating function, distribution and density of the polynomial zeros, recursion relation, asymptotics, differential or difference equations, etc.

To establish a connection with the standard formulation of quantum mechanics, we try here to develop a procedure to reconstruct the potential function in the new formulation. There is no a priori guarantee that such a procedure will always yield a potential function. However, if it works then it should be interesting especially if the derived potentials are associated with systems that do not belong to the conventional class of integrable systems in standard quantum mechanics. Examples of such systems were introduced in Refs. [1-3] where the energy polynomials are the Meixner-Pollaczek, continuous dual Hahn and Wilson polynomials in addition to their discrete versions of the Meixner, dual Hahn and Racah polynomials, respectively. In section 2, we outline the procedure of how to obtain the matrix elements of the potential in the chosen basis set $\left\{\phi_n(x)\right\}$. Then, in section 3, we present four methods of how to construct the potential function numerically using knowledge of only its computed matrix elements and the basis set. In section 4, we test the procedure by applying it first on systems with known potential functions then in section 5 we give examples where we recover potentials that do not belong to the known class of exactly solvable problems in standard quantum mechanics. Some of these potentials are recovered analytically while others can only be reconstructed numerically. Mandating the correspondence (i.e., demanding the existence of an analytic potential function) places a severe restriction on the kinematics of such problems.

## 2. The potential matrix elements

The first step is to calculate the matrix elements of the Hamiltonian operator *H*. Now, the explicit construction of the total wavefunction of the system as $\Psi(t,x) = e^{-iEt/\hbar}\psi(E,x)$ means that $H\Psi = i\hbar\frac{\partial}{\partial t}\Psi = E\Psi$. Therefore, writing the wave equation as $H|\psi\rangle = E|\psi\rangle$, where the wavefunction (1) is written as $|\psi\rangle = \sqrt{\rho^{\mu}}\sum_n P_n^{\mu}|\phi_n\rangle$, and projecting from left by $\langle\phi_n|$ gives the equation $\sum_m P_m^{\mu}\langle\phi_n|H|\phi_m\rangle = E\sum_m P_m^{\mu}\langle\phi_n|\phi_m\rangle$, which could be rewritten as the generalized eigenvalue matrix equation

$$\mathcal{H}|P\rangle = E\Omega|P\rangle, \tag{5}$$

where $\mathcal{H}$ is the matrix representation of the Hamiltonian operator in the basis $\{\phi_n\}$ and $\Omega$ is the overlap matrix of the basis elements, $\Omega_{n,m} = \langle\phi_n|\phi_m\rangle$ (i.e., matrix representation of the identity). On the other hand, the three-term recursion relation (2) could be rewritten in matrix form as $\Sigma|P\rangle = \varepsilon|P\rangle$, where $\Sigma$ is the tridiagonal symmetric matrix obtained from (2) as

$$\Sigma_{n,m} = a_n^{\mu}\delta_{n,m} + b_{n-1}^{\mu}\delta_{n,m+1} + b_n^{\mu}\delta_{n,m-1}. \tag{6}$$

Therefore, the wave equation (5) must be equivalent to and should yield the three-term recursion relation of the energy polynomials $\{P_n^{\mu}\}$. This implies that wave operator matrix $\mathcal{J} = \mathcal{H} - E\Omega$ should be tridiagonal and symmetric. This requirement places a restriction on the physical systems vis-à-vis realization of potential functions that are needed to establish a correspondence with the standard formulation. In this work, we use a notation where physical operators like the Hamiltonian and the potential are designated by Roman letters (e.g., *H* and *V*) while their matrix representation are designated by Euclid letters (e.g., $\mathcal{H}$ and $\mathcal{V}$).

In standard quantum mechanics, the Hamiltonian *H* is the sum of the kinetic energy operator *T* and the potential function *V*. Thus, $V = H - T$ and to obtain the matrix representation of the potential in the basis $\{\phi_n\}$ we need to compute the matrix elements of the kinetic energy in this basis. Now, *T* is usually a well-known differential operator in configuration space. For example, in one dimension with coordinate *x*, $T = -\frac{1}{2}\frac{d^2}{dx^2}$ and in three dimensions with spherical symmetry and radial coordinate *r*, $T = -\frac{1}{2}\frac{d^2}{dr^2} + \frac{\ell(\ell+1)}{2r^2}$, where $\ell$ is the angular momentum quantum number. Therefore, the action of *T* on the given basis elements $\{\phi_n\}$ could be derived and its matrix representation $\mathcal{T}$ depends only on the choice of basis. The matrix wave operator becomes $\mathcal{J} = \mathcal{V} + \mathcal{T} - E\Omega$, which is required to be tridiagonal since the matrix wave equation $\mathcal{J}|P\rangle = 0$ is equivalent to and should yield the three-term recursion relation $(\Sigma - \varepsilon)|P\rangle = 0$. Now, if $\Omega$ is non-tridiagonal (i.e., the basis elements are neither orthogonal nor tri-thogonal), then the kinetic energy matrix $\mathcal{T}$ must have corresponding energy-dependent components to cancel those out. This cannot be required of the potential matrix $\mathcal{V}$ since the potential is assumed to be energy independent. If after that, the matrix $\mathcal{T}$ still has some non-tridiagonal components then those must be eliminated by counter components in $\mathcal{V}$ so that the net wave operator matrix $\mathcal{J}$ becomes tridiagonal. Therefore, establishing a correspondence of our formulation with the standard potential quantum mechanics places a severe restriction on the choice of basis $\{\phi_n\}$ and potential functions to only those that result in a tridiagonal matrix representation for the wave operator $\mathcal{J} = \mathcal{V} + \mathcal{T} - E\Omega$.

As an illustration of how to obtain the kinetic energy matrix, let us consider the following square integrable basis elements in three dimensions with spherical symmetry

$$\phi_n(r) = \sqrt{\frac{\Gamma(n+1)}{\Gamma(n+2\ell+2)}}\,(\lambda r)^{\ell+1} e^{-\lambda r/2} L_n^{2\ell+1}(\lambda r), \tag{7}$$

where $L_n^{\nu}(z)$ is the Laguerre polynomial of degree $n$ in $z$ and $\lambda^{-1}$ is a length scale parameter. Using the recursion relation and orthogonality of the Laguerre polynomials, we obtain

$$\Omega_{n,m} = 2(n+\ell+1)\,\delta_{n,m} - \sqrt{n(n+2\ell+1)}\,\delta_{n,m+1} - \sqrt{(n+1)(n+2\ell+2)}\,\delta_{n,m-1}, \tag{8}$$

Moreover, the use of the differential equation of the Laguerre polynomial as well as its recursion relation and orthogonality we obtain

$$\begin{gathered}\mathcal{T}_{n,m} = -\frac{1}{2}\langle\phi_n|\left[\frac{d^2}{dr^2} - \frac{\ell(\ell+1)}{r^2}\right]|\phi_m\rangle = \\ \frac{\lambda^2}{4}\left[(n+\ell+1)\delta_{n,m} + \tfrac{1}{2}\sqrt{n(n+2\ell+1)}\delta_{n,m+1} + \tfrac{1}{2}\sqrt{(n+1)(n+2\ell+2)}\delta_{n,m-1}\right]\end{gathered} \tag{9}$$

Therefore, the two matrices $\mathcal{T}$ and $\Omega$ are compatible with the correspondence procedure since they are tridiagonal. On the other hand, let us consider the following orthonormal set of elements as basis functions in one dimension

$$\phi_n(x) = [\sqrt{\pi}\,2^n n!]^{-1/2} e^{-\lambda^2 x^2/2} H_n(\lambda x), \tag{10}$$

where $H_n(z)$ is the Hermite polynomial of degree $n$ in $z$. Using the differential equation of the Hermite polynomial, we obtain

$$\mathcal{T}_{n,m} = -\frac{1}{2}\langle\phi_n|\frac{d^2}{dx^2}|\phi_m\rangle = \lambda^2\left(n+\tfrac{1}{2}\right)\delta_{n,m} - \tfrac{1}{2}\lambda^2\langle n|(\lambda x)^2|m\rangle, \tag{11}$$

where $\langle n|f(z)|m\rangle = [\pi 2^{n+m} n!m!]^{-1/2}\int_{-\infty}^{+\infty} e^{-z^2} f(z) H_n(z) H_m(z)\,dz$. Therefore, the recursion relation of the Hermite polynomial and its orthogonality property show that the last term in (11) will produce non-tridiagonal components. In fact, this matrix component is penta-diagonal. Hence, it must be eliminated by the corresponding term $+\frac{1}{2}\lambda^4 x^2$ in the potential function, which is a harmonic oscillator potential term. This makes the sought after potential function equal to $V(x) = \frac{1}{2}\lambda^4 x^2 + \tilde{V}(x)$ and turns our task into the construction of the potential component $\tilde{V}(x)$ associated with the component $\lambda^2\left(n+\frac{1}{2}\right)\delta_{n,m}$ of the kinetic energy matrix. On the other hand, using the recursion relation of the Hermite polynomial and its orthogonality property in (11) we obtain

$$\mathcal{T}_{n,m} = -\frac{1}{2}\langle\phi_n|\frac{d^2}{dx^2}|\phi_m\rangle = \frac{\lambda^2}{4}\left[(2n+1)\,\delta_{n,m} - \sqrt{n(n-1)}\delta_{n,m+2} - \sqrt{(n+1)(n+2)}\delta_{n,m-2}\right], \tag{12}$$

which is tridiagonal in function space with only odd or only even indices. That is, if the 1D problem conserves parity, then we can split function space into two disconnected subspaces (one odd and the other even) and $\mathcal{T}_{2n,2m}$ is tridiagonal and so is $\mathcal{T}_{2n+1,2m+1}$.

Our task now is to use knowledge of the matrix elements of the potential, which is obtained as $\mathcal{V}_{n,m}=\mathcal{H}_{n,m}-\mathcal{T}_{n,m}$, and the basis in which they are computed to derive a realization of the potential function in configuration space. Sometimes, we can recover the potential function analytically. However, in most cases reconstruction of the potential function from its matrix elements and the basis can only be done numerically. To achieve the latter, we present in the following section four alternative methods with different accuracy and range of applicability. Usually, the potential matrix obtained is finite in size. Thus, the derived potential function is an approximation that should improve with an increase in the size of the potential matrix. We will demonstrate this by descriptive examples where the potential function is known a priori.

## 3. The potential function

Let $\left\{\mathcal{V}_{nm}\right\}_{n,m=0}^{N-1}$ be the $N\times N$ matrix elements of the potential function $V(x)$ in a given basis set $\left\{\phi_n(x)\right\}$ and let $\left\{\bar{\phi}_n(x)\right\}$ be the conjugate basis set; that is $\left\langle\bar{\phi}_n\middle|\phi_m\right\rangle=\left\langle\phi_n\middle|\bar{\phi}_m\right\rangle=\delta_{nm}$. Specifically, we mean that $\mathcal{V}_{nm}=\left\langle\phi_n\middle|V\middle|\phi_m\right\rangle$. Now, our task is to construct the potential function $V(x)$ using knowledge of *only* these matrix elements and the basis. This is a central issue in the classic inverse scattering problem that gives these matrix elements as the final product. Here, it is also necessary to establish the correspondence of our formulation of quantum mechanics, where the potential function is not an ingredient, with the conventional formulation in which the potential function is an essential component of the formulation.

### 3.1 First method:

Let $V$ denote the quantum mechanical Hermitian operator that stands for the real potential energy. Then using Dirac notation, we can write

$$\left\langle x\middle|V\middle|x'\right\rangle=V(x)\delta(x-x'), \tag{13}$$

where $\delta(x-x')=\left\langle x\middle|x'\right\rangle$ and where $x$ stands for the configuration space coordinate. Moreover, $\left\langle x\middle|\phi_n\right\rangle=\phi_n(x)$ and $\left\langle x\middle|\bar{\phi}_n\right\rangle=\bar{\phi}_n(x)$. The completeness of the basis reads as follows $\sum_n\left|\bar{\phi}_n\right\rangle\left\langle\phi_n\right|=\sum_n\left|\phi_n\right\rangle\left\langle\bar{\phi}_n\right|=I$, where $I$ is the identity. It enables us to rewrite the left side of (13) as

$$\begin{aligned}\sum_{n,m=0}^{\infty}\left\langle x\middle|\bar{\phi}_n\right\rangle\left\langle\phi_n\middle|V\middle|\phi_m\right\rangle\left\langle\bar{\phi}_m\middle|x'\right\rangle&=\sum_{n,m=0}^{\infty}\bar{\phi}_n(x)\mathcal{V}_{nm}\bar{\phi}_m(x')\\&=\frac{1}{2}\sum_{n,m=0}^{\infty}\mathcal{V}_{nm}\left[\bar{\phi}_n(x)\bar{\phi}_m(x')+\bar{\phi}_m(x)\bar{\phi}_n(x')\right]\end{aligned}\tag{14}$$

where we have assumed reality of the basis elements such that $\left\langle\bar{\phi}_m\middle|x'\right\rangle=\left\langle x'\middle|\bar{\phi}_m\right\rangle$. The right side of Eq. (13) reads

$$V(x)\sum_{n=0}^{\infty}\left\langle x\right|\frac{\left|\phi_n\right\rangle\left\langle\bar{\phi}_n\right|+\left|\bar{\phi}_n\right\rangle\left\langle\phi_n\right|}{2}\left|x'\right\rangle=\frac{1}{2}V(x)\sum_{n=0}^{\infty}\left[\phi_n(x)\bar{\phi}_n(x')+\bar{\phi}_n(x)\phi_n(x')\right]. \tag{15}$$

Therefore, we can write the approximation

$$V(x)\sum_{n=0}^{N-1}\left[\phi_n(x)\bar{\phi}_n(x')+\bar{\phi}_n(x)\phi_n(x')\right]\cong\sum_{n,m=0}^{N-1}\mathcal{V}_{nm}\left[\bar{\phi}_n(x)\bar{\phi}_m(x')+\bar{\phi}_m(x)\bar{\phi}_n(x')\right], \tag{16}$$

where $N$ is some large enough natural number. Taking $x'=x$, gives

$$V(x)\cong\frac{\sum_{n,m=0}^{N-1}\mathcal{V}_{nm}\overline{\phi}_n(x)\overline{\phi}_m(x)}{\sum_{n=0}^{N-1}\phi_n(x)\overline{\phi}_n(x)}\,. \tag{17}$$

Of course, this is an approximation which is valid for finite values of $N$ since the denominator blows up in the limit as $N\to\infty$ because $\sum_{n=0}^{\infty}\phi_n(x)\overline{\phi}_n(x')=\sum_{n=0}^{\infty}\overline{\phi}_n(x)\phi_n(x')=\delta(x-x')$.

### 3.2 Second method:

Using the completeness of (or the identity in) configuration space, $\int|x'\rangle\langle x'|dx'=1$, and Eq. (13) we can write

$$\langle x|V|\phi_n\rangle=\int\langle x|V|x'\rangle\langle x'|\phi_n\rangle dx'=V(x)\phi_n(x)\,. \tag{18}$$

The completeness of the basis enables us to write the left side of this equation as

$$\langle x|V|\phi_n\rangle=\sum_{m=0}^{\infty}\langle x|\overline{\phi}_m\rangle\langle\phi_m|V|\phi_n\rangle=\sum_{m=0}^{\infty}\overline{\phi}_m(x)\mathcal{V}_{mn}\cong\sum_{m=0}^{N-1}\overline{\phi}_m(x)\mathcal{V}_{mn}\,. \tag{19}$$

These two equations give

$$V(x)\cong\sum_{m=0}^{N-1}\frac{\overline{\phi}_m(x)}{\phi_n(x)}\mathcal{V}_{mn}\,,\quad n=0,1,...,N-1 \tag{20}$$

Therefore, we need the information in only one column of the potential matrix (or one row, since $\mathcal{V}_{mn}=\mathcal{V}_{nm}$) to determine $V(x)$. In particular, if we choose $n = 0$, we obtain

$$V(x)\cong\sum_{m=0}^{N-1}\frac{\overline{\phi}_m(x)}{\phi_0(x)}\mathcal{V}_{m,0}\,. \tag{21}$$

When comparing this method with the other three, we should take into consideration that this method uses only $N$ elements of the potential matrix while the others use $\frac{1}{2}N(N+1)$ .

### 3.3 Third method:

Completeness of the basis enables us to integrate (13) over $x'$ as follows

$$\begin{aligned}V(x)&=\int\langle x|V|x'\rangle dx'=\sum_{n,m=0}^{\infty}\int\langle x|\overline{\phi}_n\rangle\langle\phi_n|V|\phi_m\rangle\langle\overline{\phi}_m|x'\rangle dx'\\&=\sum_{n,m=0}^{\infty}\int\overline{\phi}_n(x)\mathcal{V}_{nm}\overline{\phi}_m(x')dx'\cong\sum_{n,m=0}^{N-1}\mathcal{V}_{nm}\overline{\phi}_n(x)\int\overline{\phi}_m(x')dx'\end{aligned} \tag{22}$$

The integral in the last equation could be evaluated using any appropriate scheme. For example, we can use a Gauss quadrature that is compatible with the $L^2$ space spanned by $\{\overline{\phi}_n(x)\}$ (see Appendix A for details):

$$\int\overline{\phi}_m(x')dx'=\int\rho(x')\frac{\overline{\phi}_m(x')}{\rho(x')}dx'=\sum_{k=0}^{K-1}\omega_k\frac{\overline{\phi}_m(\tau_k)}{\rho(\tau_k)}=\sum_{k=0}^{K-1}\omega_k^d\overline{\phi}_m(\tau_k)\,, \tag{23}$$

for some large enough integer $K \geq N$ and $\{\tau_k, \omega_k\}_{k=0}^{K-1}$ are the abscissa and numerical weights of the quadrature, while the *derivative weights* $\{\omega_k^d\}_{k=0}^{K-1}$ are $\omega_k^d = \omega_k / \rho(\varepsilon_k)$. Therefore, we can write equation (22) as

$$V(x) \cong \sum_{n,m=0}^{N-1} \left[ \sum_{k=0}^{K-1} \mathcal{V}_{nm} \omega_k^d \bar{\phi}_m(\tau_k) \right] \bar{\phi}_n(x) \equiv \sum_{n=0}^{N-1} U_n \bar{\phi}_n(x), \tag{24}$$

where

$$U_n = \sum_{m=0}^{N-1} \sum_{k=0}^{K-1} \mathcal{V}_{nm} \omega_k^d \bar{\phi}_m(\tau_k). \tag{25}$$

Thus, in this method one has to make an appropriate choice of integration scheme, which in our case is reflected by the abscissa and numerical weights of the quadrature, $\{\tau_k, \omega_k\}_{k=0}^{K-1}$. In the last paragraph of Appendix A, we give a representation of $\{U_n\}$ in a special basis like those given by (7) and (10) above.

**3.4 Fourth method:**

Let $\{p_n(x)\}$ be a complete set of orthogonal polynomials with weight function $\rho(x)$. If the $L^2$ basis elements are chosen as $\phi_n(r) = \sqrt{\gamma(x)\rho(x)}\, p_n(x)$ where $x = x(r)$ and $\gamma(x)$ is some proper function (see Appendix A for details), then there is a simple way to obtain $V(r)$ from its matrix elements $\{\mathcal{V}_{nm}\}_{n,m=0}^{N-1}$ in this basis, which is computed using the Gauss quadrature associated with the orthogonal polynomials $\{p_n(x)\}$ as shown in the Appendix and as follows

$$\begin{aligned} \mathcal{V}_{nm} &= \langle \phi_n | V | \phi_m \rangle = \lambda \int \phi_n(r) V(r) \phi_n(r) dr = \int \rho(x)\gamma(x) V\left(r(x)\right) p_n(x) p_m(x) \frac{dx}{x'} \\ &\cong \sum_{k=0}^{N-1} \omega_k \left[ \gamma(\tau_k) V\left(r(\tau_k)\right) / x'(\tau_k) \right] p_n(\tau_k) p_m(\tau_k) = \sum_{k=0}^{N-1} \left[ \gamma(\tau_k) V\left(r(\tau_k)\right) / x'(\tau_k) \right] \Lambda_{nk} \Lambda_{mk} \end{aligned} \tag{26}$$

where $x'(x) = \lambda^{-1}(dx/dr)$. In matrix notation, this equation could be written as $\mathcal{V} = \Lambda W \Lambda^{\mathrm{T}}$, where $W$ is $N \times N$ diagonal matrix with elements $W_{kk} = \gamma(\tau_k) V\left(r(\tau_k)\right) / x'(\tau_k)$. Now, since $\Lambda$ is the normalized eigenvectors matrix then $\Lambda^{\mathrm{T}}\Lambda = \Lambda\Lambda^{\mathrm{T}} = 1$ giving $W = \Lambda^{\mathrm{T}} \mathcal{V} \Lambda$. Then any suitable function fitting routine could be used to compute $V(r)$ from its $N$ values $\{W_{nn} x'(\tau_n)/\gamma(\tau_n)\}_{n=0}^{N-1}$ at the points $\{r(\tau_n)\}_{n=0}^{N-1}$. So, given the matrix elements of the potential in the special basis above, we start by calculating the eigenvalues of the tridiagonal quadrature matrix $J$ in Eq. (A3) of Appendix A and their corresponding normalized eigenvectors. Thus, we obtain the diagonal matrix $W$ that leads to $V\left(r(\tau_n)\right)$ and use a fitting routine to obtain $V(r)$. In this work, we use the continued fraction fitting routine based on the rational fraction approximation of Haymaker and Schlessinger similar to that in the Padé method [5].

To test the above four methods, we select a priori the radial potential function $V(r) = 5r^2 e^{-r}$ in 3D and compute its matrix elements in the 3D basis of Eq. (7) with $\ell = 1$ and $\lambda = 7$ using the Gauss quadrature associated with the Laguerre polynomials. Then, we reconstruct the potential function using only these matrix elements and the basis as described above. Figure 1, shows that the third and fourth methods produce the best result.

To make a stringent test of the third and fourth methods, we choose another nonanalytic potential function. Specifically, we consider the piece-wise linear potential function

$$\bar{V}(r)=\begin{cases} 2r & 0\le r<1.2 \\ 2.4 & 1.2\le r<3 \\ 4.2-0.6r & 3\le r<7 \\ 0 & \text{otherwise} \end{cases} \tag{27}$$

which is constructed to trace the function $V(r)=5r^2e^{-r}$. Figure 2 is a plot of the two potential functions: $V(r)$ and its nonanalytic version $\bar{V}(r)$. Figure 3 shows that the third method is superior to the fourth and illustrates the accuracy of the result with an increase in the size of the potential matrix. This may lead to the wrong conclusion that we should use only the third method in our study. However, in some cases (as will be demonstrated below) the other methods can produce superior results. Consequently, we will keep using all four methods in subsequent investigations.

In the following two sections, we use the procedure outlined in Sec. 2 above to compute the matrix elements of the potential associated with a given system in our formulation of quantum mechanics. Then if the potential function could not be reconstructed analytically from these matrix elements and the basis, then we use one or more of the four methods to obtain an approximate graphical representation of the potential function. To demonstrate reliability and accuracy of the procedure, we start in the following section with two systems that have corresponding ones in the standard formulation of quantum mechanics with well-known potential functions. Thereafter in section 5, we present our original findings where we reconstruct the potential function of exactly solvable problems in our formulation none of which belong to the well-known class of exactly solvable problems in the conventional formulation of quantum mechanics.

## 4. Establishing the correspondence: Conventional problems

### 4.1 The three-dimensional Coulomb problem:

This problem was treated using our reformulation of quantum mechanics in section IV.A of Ref. [2]. The energy polynomial that enters in Eq. (1) is the two-parameter Meixner-Pollaczek polynomial $P_n^{\mu}(y,\theta)$ defined in Appendix B by Eq. (B1) with $y=Z/k=Z/\sqrt{2E}$, $\mu=\ell+1$ and $\cos\theta=\frac{4k^2-\lambda^2}{4k^2+\lambda^2}$. These polynomials satisfy the symmetric three-term recursion relation shown as Eq. (B4) in the same Appendix. A proper basis for this problem is the one given by Eq. (7). Comparing the asymptotics of this polynomial, which is given by formula (B6), to the general formula (3) shows that the scattering phase shift (modulo $\pi/2$) is

$$\delta(\varepsilon)=\arg\Gamma\left(\ell+1+\mathrm{i}\,Z/k\right), \tag{28}$$

In standard quantum mechanics, this scattering phase shift is associated with the 3D Coulomb potential $V_C(r)=-Z/r$, where $Z$ is the electric charge. Nonetheless, we will use the correspondence procedure outlined above to establish this fact. Substituting the physical parameters in the three-term recursion relation (B4), we obtain

$$(\lambda Z)P_n^{\ell+1} = -2(n+\ell+1)\left(E-\tfrac{1}{8}\lambda^2\right)P_n^{\ell+1}$$
$$+\left(E+\tfrac{1}{8}\lambda^2\right)\left[\sqrt{n(n+2\ell+1)}P_{n-1}^{\ell+1}+\sqrt{(n+1)(n+2\ell+2)}P_{n+1}^{\ell+1}\right] \tag{29}$$

We rewrite this as the wave equation $\mathcal{H}|P\rangle = E\Omega|P\rangle$, where $\Omega$ is calculated using the basis (7) and given by Eq. (8) . Thus, we obtain the following matrix elements of the Hamiltonian

$$\mathcal{H}_{n,m} = \left[\tfrac{1}{4}\lambda^2(n+\ell+1)-\lambda Z\right]\delta_{n,m}$$
$$+\tfrac{1}{8}\lambda^2\left[\sqrt{n(n+2\ell+1)}\,\delta_{n,m+1}+\sqrt{(n+1)(n+2\ell+2)}\,\delta_{n,m-1}\right] \tag{30}$$

Finally, the matrix representation of the potential function in the basis (7) are obtained as $\mathcal{V} = \mathcal{H} - \mathcal{T}$, where the kinetic energy matrix $\mathcal{T}$ is given by Eq. (9). As a result, we obtain in fact a simple exact expression for the matrix elements of the potential in the appropriate basis (7) as $\mathcal{V}_{n,m} = -\lambda Z\delta_{n,m}$. Using these and the basis (7) with $\bar{\phi}_n(r) = (\lambda r)^{-1}\phi_n(r)$, we obtain an approximation of the potential function using the four methods given in the previous section. In this case, and for this potential, all methods except the third produce an exact match with the Coulomb potential $V_C(r)$ for any basis size *N*. Nonetheless, we give in Figure 4 the result obtained by the third method that improves with increasing *N*. It is interesting to note that the simple identity matrix representation $\delta_{n,m}$ in the basis (7) is obtained as matrix elements of the function $(\lambda r)^{-1}$ in the basis (7). That is, $\langle\phi_n|(\lambda r)^{-1}|\phi_m\rangle = \delta_{n,m}$ giving the analytic realization of the potential as $-Z/r$.

### 4.2 The one-dimensional Morse problem:

This problem was also treated using our formulation in section IV.B of Ref. [2] where the energy polynomial that enters in Eq. (1) is the three-parameter continuous dual Hahn polynomial $S_n^\mu(y^2;a,b)$ defined by Eq. (B9) in Appendix B with $y = k/\lambda = \lambda^{-1}\sqrt{2E}$, $a = b = \alpha+1$ and $\mu = \frac{1}{2} - 4\beta\left(V_0/\lambda^2\right)$ where $V_0$ is a real parameter of inverse square length, $\alpha$ and $\beta$ are a dimensionless parameters. This polynomial satisfies the symmetric three-term recursion relation given by Eq. (B11). A proper basis for this problem is the following orthonormal set

$$\phi_n(x) = \sqrt{\tfrac{\Gamma(n+1)}{\Gamma(n+2\alpha+2)}}\, z^{\alpha+1}e^{-z/2}L_n^{2\alpha+1}(z), \tag{31}$$

where $z = e^{\lambda x}$ with $-\infty < x < +\infty$ and we require that $\alpha > -1$. From the asymptotics of this polynomial, which is given by Eq. (B13), we obtain the scattering phase shift as

$$\delta(\varepsilon) = \arg\Gamma\left(2i\lambda^{-1}k\right) - 2\arg\Gamma\left(\alpha+1+i\lambda^{-1}k\right) - \arg\Gamma\left(\tfrac{1}{2}-\tfrac{4\beta}{\lambda^2}V_0+i\lambda^{-1}k\right). \tag{32}$$

In standard quantum mechanics, this scattering phase shift is associated with the 1D Morse potential $V_M(x) = V_0\left(e^{2\lambda x}-2\beta e^{\lambda x}\right)$ [6]. Substituting the physical quantities in the three-term recursion relation (B11) we obtain the matrix equation $E|P\rangle = (\lambda^2/2)\Sigma|P\rangle$, where the matrix elements of $\Sigma$ is obtained as

$$\Sigma_{n,n} = (n+\mu+\alpha+1)^2 + \left(n+\alpha+\tfrac{1}{2}\right)^2 - \mu^2 - \left(\alpha+\tfrac{1}{2}\right)^2$$
$$\Sigma_{n,n+1} = \Sigma_{n+1,n} = -(n+\mu+\alpha+1)\sqrt{(n+1)(n+2\alpha+2)} \tag{33}$$

On the other hand, the matrix wave equation (5) could be rewritten as $E|P\rangle = \Omega^{-1}\mathcal{H}|P\rangle = \mathcal{H}|P\rangle$ since $\Omega_{n,m} = \langle\phi_n|\phi_m\rangle = \delta_{n,m}$. Hence, we conclude that the Hamiltonian matrix is $\mathcal{H} = (\lambda^2/2)\Sigma$. Finally, we obtain the matrix elements of the potential function in the basis (31) as $\mathcal{V} = \mathcal{H} - \mathcal{T}$, where the elements of the matrix $\mathcal{T}$ are $\mathcal{T}_{n,m} = -\frac{1}{2}\langle\phi_n|\frac{d^2}{dx^2}|\phi_m\rangle$. Using $\frac{d}{dx} = \lambda z\frac{d}{dz}$ and the differential equation of the Laguerre polynomial together with its differential property, $z\frac{d}{dz}L_n^\nu(z) = nL_n^\nu(z) - (n+\nu)L_{n-1}^\nu(z)$, we obtain

$$-\frac{2}{\lambda^2}\mathcal{T}_{n,m} = \left[n + (\alpha+1)^2\right]\delta_{n,m} - \sqrt{n(n+2\alpha+1)}\,\delta_{n,m+1} + \tfrac{1}{4}\langle n|z^2|m\rangle - \tfrac{1}{2}(2n+2\alpha+3)\langle n|z|m\rangle \tag{34}$$

where $\langle n|f(z)|m\rangle = \sqrt{\frac{\Gamma(n+1)\Gamma(m+1)}{\Gamma(n+\nu+1)\Gamma(m+\nu+1)}}\int_0^\infty z^\nu e^{-z} f(z) L_n^\nu(z) L_m^\nu(z)\,dz$. The recursion relation of the Laguerre polynomials and their orthogonality show that the third term on the right side of Eq. (34) produce non-tridiagonal matrix elements. Thus, the potential function must contain a counter term to cancel it so that the non-tridiagonal matrix component be eliminated. That is, the potential function must contain the term $+\frac{1}{8}\lambda^2 z^2 = \frac{1}{8}\lambda^2 e^{2\lambda x}$ giving $V(x) = \frac{1}{8}\lambda^2 e^{2\lambda x} + \tilde{V}(x)$. Thus, our search will then be for the component $\tilde{V}(x)$ which is associated with the kinetic energy matrix (34) without the $\frac{1}{4}\langle n|z^2|m\rangle$ term and which reads as follows

$$-\frac{2}{\lambda^2}\tilde{\mathcal{T}}_{n,m} = \left[n + (\alpha+1)^2\right]\delta_{n,m} - \sqrt{n(n+2\alpha+1)}\,\delta_{n,m+1} - \tfrac{1}{2}(2n+2\alpha+3)\langle n|z|m\rangle. \tag{35}$$

To compute the last term, we use the recursion relation of the Laguerre polynomials and their orthogonality property while noting that $\lambda\int_{-\infty}^{+\infty} \ldots\, dx = \int_0^\infty \ldots\, z^{-1}dz$. As a result, we obtain

$$-\frac{2}{\lambda^2}\tilde{\mathcal{T}}_{n,m} = \left[\alpha(\alpha+1) - 2(n+\alpha+1)^2\right]\delta_{n,m} + \tfrac{1}{2}(2n+2\alpha+3)\sqrt{(n+1)(n+2\alpha+2)}\,\delta_{n,m-1} + \tfrac{1}{2}(2n+2\alpha+1)\sqrt{n(n+2\alpha+1)}\,\delta_{n,m+1} \tag{36}$$

Using the Hamiltonian matrix from (33) and this kinetic energy matrix we obtain the following potential matrix corresponding to $\tilde{V}(x)$ as $(\lambda^2/2)\Sigma - \tilde{\mathcal{T}}$ giving

$$\frac{4}{\lambda^2}\tilde{\mathcal{V}}_{n,m} = (2\mu-1)\left[2(n+\alpha+1)\delta_{n,m} - \sqrt{n(n+2\alpha+1)}\,\delta_{n,m+1} - \sqrt{(n+1)(n+2\alpha+2)}\,\delta_{n,m-1}\right] \tag{37}$$

Now, using these matrix elements and the basis (31), we calculate the potential function component $\tilde{V}(x)$. For this problem, the second method produces an exact match to $V_M(x)$ with $V_0 = \frac{1}{8}\lambda^2$ for any basis size *N*. The third method produces only oscillations that diverge with increasing *N*. Figure 5 shows the results of the first and fourth methods. On the other hand, we can show that the three terms inside the square brackets of the potential matrix (37) is obtained as $\langle\phi_n|z|\phi_m\rangle$. Thus, we can write the exact realization $\tilde{V}(x) = \frac{\lambda^2}{4}(2\mu-1)z = -2\beta V_0 e^{\lambda x}$ giving the full potential function as $V(x) = \frac{1}{8}\lambda^2 e^{2\lambda x} + \tilde{V}(x) = \frac{1}{8}\lambda^2(e^{2\lambda x} - 2\beta e^{\lambda x})$.

With these two illustrative examples, we demonstrated that:

- Our strategy for establishing the correspondence between our reformulation of quantum mechanics without a potential and the standard formulation works since we were able to reconstruct the potential function either analytically or numerically.
- All four numerical methods established above should be used in the process since their accuracy differ from one problem to the other.

So far, we established correspondence between our formulation of quantum mechanics without a potential and the standard formulation by reconstructing the potential functions associated with well-known sample problems that included the Coulomb and Morse potentials. Next, we reconstruct the potential functions for problems that do not belong to the conventional class of exactly solvable potentials. We divide the problems into two classes. One where we could reconstruct the potential function analytically whereas for the other we could do that only numerically.

## 5. Establishing the correspondence: Nonconventional problems

### 5.1 Analytic realization of the potential

In this subsection, we give two examples to demonstrate how to recover the potential function analytically. We take the following square integrable Jacobi basis

$$\phi_n(x) = A_n(1-z^2)^\alpha P_n^{(\nu,\nu)}(z), \tag{38}$$

where $P_n^{(\mu,\nu)}(z)$ is the Jacobi polynomial of degree *n* in *z* and the real parameters are such that $\alpha > 0$ and $\nu > -1$. The coordinate transformation *z*(*x*) is such that $-1 \le z \le +1$ and the normalization constant is $A_n = \frac{1}{\Gamma(n+\nu+1)}\sqrt{\frac{2n+2\nu+1}{2^{2\nu+1}}\Gamma(n+1)\Gamma(n+2\nu+1)}$. We consider two cases in the following two subsections. Both do not belong to the conventional class of exactly solvable problems in the standard formulation of quantum mechanics. However, we will be able to give in both cases an analytic reconstruction of the potential function

#### 5.1.1 *Sinusoidal potential box*:

We take $z(x) = \sin(\lambda x)$ where $-\pi/2\lambda \le x \le +\pi/2\lambda$. Then, we can show that if $2\alpha = \nu + \frac{1}{2}$ then $\Omega$ becomes diagonal (i.e., the basis elements (38) are orthogonal) and the kinetic energy matrix reads as follows

$$\frac{2}{\lambda^2}\mathcal{T}_{n,m} = \left(n+\nu+\tfrac{1}{2}\right)^2 \delta_{n,m} - \left(\nu^2 - \tfrac{1}{4}\right)\langle n|\frac{1}{1-z^2}|m\rangle, \tag{39}$$

where

$$\langle n|\,f(z)\,|m\rangle = A_n A_m \int_{-1}^{+1}(1-z^2)^\nu\, f(z) P_n^{(\nu,\nu)}(z) P_m^{(\nu,\nu)}(z)dz. \tag{40}$$

The recursion relation of the Jacobi polynomials and their orthogonality show that the second term on the right side of Eq. (39) produces non-tridiagonal matrix elements. Thus, the potential function must contain a counter term to cancel it so that we end up with only a tridiagonal matrix. That is, the potential function must contain the term $\frac{\lambda^2}{2}\frac{\nu^2-\frac{1}{4}}{1-z^2} = \frac{\lambda^2}{2}\frac{\nu^2-\frac{1}{4}}{\cos^2(\lambda x)}$ giving $V(x) = \frac{V_2}{\cos^2(\lambda x)} + \tilde{V}(x)$ with the basis parameter $\nu^2 = \frac{1}{4} + 2V_2/\lambda^2$ requiring the potential parameter $V_2 \geq -\lambda^2/8$. Thus, our search will be for the component $\tilde{V}(x)$ which is associated with the kinetic energy matrix

$$\tilde{\mathcal{T}}_{n,m} = \frac{\lambda^2}{2}\left(n + \nu + \tfrac{1}{2}\right)^2 \delta_{n,m}. \tag{41}$$

Now, since $\Omega$ and $\tilde{\mathcal{T}}$ are diagonal then to maintain the tridiagonal structure of the wave operator we require that the matrix elements $\langle \phi_n | \tilde{V}(x) | \phi_m \rangle$ must be at most tridiagonal. Noting that $\lambda \int_{-\pi/2\lambda}^{+\pi/2\lambda} ... \, dx = \int_{-1}^{+1} ... \frac{dz}{\sqrt{1-z^2}}$, then $\langle \phi_n | \tilde{V}(x) | \phi_m \rangle = \langle n | \tilde{V}(x) | m \rangle$ and the recursion relation of the Jacobi polynomial and its orthogonality dictate that $\tilde{V}(x)$ be a linear function in *z*. That is, $\tilde{V}(x) = V_0 + V_1 \sin(\lambda x)$ and then the total potential function becomes

$$V(x) = V_0 + V_1 \sin(\lambda x) + \frac{V_2}{\cos^2(\lambda x)}, \tag{42}$$

where $\nu^2 = \frac{1}{4} + \frac{2V_2}{\lambda^2}$. Figure 6 is a plot of this potential obtained by varying one parameter while keeping the other two fixed. The first and last term of the potential correspond to well-known and exactly solvable potentials in standard quantum mechanics. They constitute a special case of either the trigonometric Pöschl-Teller potential or the trigonometric Scarf potential. Nonetheless, with $V_1 \neq 0$ this potential, which is exactly solvable in our reformulation, does not belong to the conventional class of exactly solvable problems. The pure $V_1$ potential term represents a potential box with sinusoidal bottom, which is not known to have an exact solution in standard quantum mechanics [7]. Using the recursion relation of the Jacobi polynomial and its orthogonality, we obtain the following matrix elements of the potential function $\tilde{V}(x)$

$$\tilde{\mathcal{V}}_{n,m} = V_0 \delta_{n,m} + \tfrac{1}{2} V_1 \left[ \sqrt{\frac{n(n+2\nu)}{(n+\nu)^2 - 1/4}}\, \delta_{n,m+1} + \sqrt{\frac{(n+1)(n+2\nu+1)}{(n+\nu+1)^2 - 1/4}}\, \delta_{n,m-1} \right]. \tag{43}$$

Adding to this the corresponding kinetic energy matrix $\tilde{\mathcal{T}}$ of Eq. (41) we obtain the Hamiltonian matrix. With $\Omega_{n,m} = \langle \phi_n | \phi_m \rangle = \delta_{n,m}$, the wave equation (5), $\mathcal{H}|P\rangle = E\Omega|P\rangle$, results in the following symmetric three-term recursion relation for the corresponding energy polynomial

$$\varepsilon P_n(\varepsilon) = \left[\left(n + \nu + \tfrac{1}{2}\right)^2 + u_0\right] P_n(\varepsilon) + \tfrac{1}{2} u_1 \left[ \sqrt{\frac{n(n+2\nu)}{(n+\nu)^2 - \frac{1}{4}}}\, P_{n-1}(\varepsilon) + \sqrt{\frac{(n+1)(n+2\nu+1)}{(n+\nu+1)^2 - \frac{1}{4}}}\, P_{n+1}(\varepsilon) \right] \tag{44}$$

where $\varepsilon = 2E/\lambda^2$ and $u_i = 2V_i/\lambda^2$. This energy polynomial and its corresponding weight function are those that enter in the expansion series of the wavefunction (1) for this problem.

### 5.1.2 *Hyperbolic potential pulse*:

We take $z(x) = \tanh(\lambda x)$, where $-\infty < x < +\infty$, and the basis parameters $2\alpha = \nu$. Noting that $\lambda \int_{-\infty}^{+\infty} ... \, dx = \int_{-1}^{+1} ... \frac{dz}{1-z^2}$, then the basis overlap matrix becomes $\Omega_{n,m} = \langle \phi_n | \phi_m \rangle = \langle n | (1-z^2)^{-1} | m \rangle$ which is non-tridiagonal but, in fact, a full matrix. As stated in section 2 above, the non-tridiagonal component must be eliminated by an energy dependent counter component in the kinetic energy matrix, which is obtained as

$$\mathcal{T}_{n,m}=\frac{\lambda^2}{2}\left\{\left[\left(n+\nu+\tfrac{1}{2}\right)^2-\tfrac{1}{4}\right]\delta_{n,m}-\nu^2\langle n|\frac{1}{1-z^2}|m\rangle\right\}. \tag{45}$$

Therefore, to cancel the non-tridiagonal terms in $\Omega$, we require that $\nu^2=-2E/\lambda^2$ making the basis parameter $\nu$ energy dependent and dictating that an exact solution is obtained only for negative energy. Moreover, we obtain the following wave operator matrix

$$\mathcal{J}_{n,m}=\mathcal{V}_{n,m}+\mathcal{T}_{n,m}-E\Omega_{n,m}=\langle n|\frac{V(x)}{1-z^2}|m\rangle+\frac{\lambda^2}{2}\left[\left(n+\nu+\tfrac{1}{2}\right)^2-\tfrac{1}{4}\right]\delta_{n,m}. \tag{46}$$

Therefore, the tridiagonal requirement of this matrix and the recursion relation of the Jacobi polynomial and its orthogonality dictate that $\frac{V(x)}{1-z^2}=V_0+V_1z$ giving the following potential function

$$V(x)=\frac{V_0+V_1\tanh(\lambda x)}{\cosh^2(\lambda x)}. \tag{47}$$

With $V_1\neq 0$, this potential does not belong to the conventional class of exactly solvable problems. Figure 7 is a plot of this potential obtained by varying $V_0$ while keeping $V_1$ fixed. Since the solution is valid for negative energy then the ratio $|V_0/V_1|$ must be less than one [8]. Finally, the wave operator matrix (46) becomes

$$\frac{2}{\lambda^2}\mathcal{J}_{n,m}=\left[\left(n+\nu+\tfrac{1}{2}\right)^2-\tfrac{1}{4}+u_0\right]\delta_{n,m}+\tfrac{1}{2}u_1\left[\sqrt{\tfrac{n(n+2\nu)}{(n+\nu)^2-1/4}}\,\delta_{n,m+1}+\sqrt{\tfrac{(n+1)(n+2\nu+1)}{(n+\nu+1)^2-1/4}}\,\delta_{n,m-1}\right] \tag{48}$$

where $u_i=2V_i/\lambda^2$. Note that the energy is embedded in the parameter $\nu$. The matrix wave equation $\mathcal{J}|P\rangle=0$ gives the following three-term recursion relation for the corresponding energy polynomials that enter as expansion coefficients of the wave function (1)

$$-u_0P_n(\varepsilon)=\left[\left(n+\nu+\tfrac{1}{2}\right)^2-\tfrac{1}{4}\right]P_n(\varepsilon)+\tfrac{1}{2}u_1\left[\sqrt{\tfrac{n(n+2\nu)}{(n+\nu)^2-\frac{1}{4}}}\,P_{n-1}(\varepsilon)+\sqrt{\tfrac{(n+1)(n+2\nu+1)}{(n+\nu+1)^2-\frac{1}{4}}}\,P_{n+1}(\varepsilon)\right] \tag{49}$$

**5.2 Numerical realization of the potential:**

In this subsection, we present two examples where we will not be able to recover the potential function analytically. Nonetheless, we can still reconstruct it numerically by using any of the four techniques established in section 3 above. Here, we consider a physical system on the positive real line where the basis elements are chosen as follows

$$\phi_n(x)=A_n(1-z)^{\frac{\mu+1}{2}}(1+z)^{\frac{\nu+1/2}{2}}P_n^{(\mu,\nu)}(z), \tag{50}$$

where $z(x)=2\tanh^2(\lambda x)-1$ and the normalization is $A_n=\sqrt{\frac{2n+\mu+\nu+1}{2^{\mu+\nu}\sqrt{2}}\frac{\Gamma(n+1)\Gamma(n+\mu+\nu+1)}{\Gamma(n+\mu+1)\Gamma(n+\nu+1)}}$. Noting that $\lambda\int_0^\infty\ldots dx=\frac{1}{\sqrt{2}}\int_{-1}^{+1}\ldots\frac{dz}{(1-z)\sqrt{1+z}}$, then we obtain the basis overlap matrix as $\Omega_{n,m}=\langle\phi_n|\phi_m\rangle=\delta_{n,m}$ meaning that the basis elements (50) form an orthonormal set. On the other hand, the kinetic energy matrix $\mathcal{T}$ in this basis has the following elements

$$\frac{2}{\lambda^2}\mathcal{T}_{n,m} = -\frac{1}{2}\left(\nu^2-\tfrac{1}{4}\right)\langle n|\frac{1-z}{1+z}|m\rangle$$
$$-\left\{\frac{2n(n+\nu)}{2n+\mu+\nu}+\frac{(\mu+1)^2}{2}+\left[\left(n+\tfrac{\mu+\nu}{2}+1\right)^2-\frac{1}{16}\right](C_n-1)\right\}\delta_{n,m}\,, \tag{51}$$
$$-\left[\left(n+\tfrac{\mu+\nu}{2}\right)^2-\frac{1}{16}\right]D_{n-1}\delta_{n,m+1}-\left[\left(n+\tfrac{\mu+\nu}{2}+1\right)^2-\frac{1}{16}\right]D_n\delta_{n,m-1}$$

where $C_n = \frac{\nu^2-\mu^2}{(2n+\mu+\nu)(2n+\mu+\nu+2)}$, $D_n = \frac{2}{2n+\mu+\nu+2}\sqrt{\frac{(n+1)(n+\mu+1)(n+\nu+1)(n+\mu+\nu+1)}{(2n+\mu+\nu+1)(2n+\mu+\nu+3)}}$ and we have used the differential equation, recursion relation and orthogonality of the Jacobi polynomial. Moreover, we have defined

$$\langle n|\,f(z)\,|m\rangle = \frac{A_nA_m}{\sqrt{2}}\int_{-1}^{+1}(1-z)^{\mu}(1+z)^{\nu}f(z)P_n^{(\mu,\nu)}(z)P_m^{(\mu,\nu)}(z)dz\,. \tag{52}$$

Now, the first term in the kinetic energy matrix (51) is not tridiagonal and has to be eliminated by a counter term from the potential function. Therefore, we write the potential as $V(x)=V_2\frac{1-z}{1+z}+\tilde{V}(x)=\frac{V_2}{\sinh^2(\lambda x)}+\tilde{V}(x)$, where $\nu^2=\frac{1}{4}+\frac{4}{\lambda^2}V_2$ and thus reality requires that the potential parameter $V_2 \geq -\left(\lambda/4\right)^2$. In the following two subsections, we obtain a numerical reconstruction of the potential component $\tilde{V}(x)$ for two different physical systems.

### 5.2.1 *Continuous dual Hahn system*:

The wavefunction for this system is written in the standard format (1) and as follows

$$\psi(E,x)=\sqrt{\rho^{\gamma}(\varepsilon;a,b)}\sum_{n=0}^{\infty}S_n^{\gamma}(\varepsilon;a,b)\phi_n(x)\,, \tag{53}$$

where $S_n^{\gamma}(\varepsilon;a,b)$ is the three-parameter continuous dual Hahn polynomial defined in Appendix B by Eq. (B9) and $\rho^{\gamma}(\varepsilon;a,b)$ is the corresponding weight function (B10). Moreover, we choose the physical parameters as $a=b=\mu+1$ and $\varepsilon=2E/\lambda^2$. Therefore, the recursion relation (B11) becomes $\varepsilon|P\rangle=\Sigma|P\rangle$, where

$$\Sigma_{n,n}=\left(n+\gamma+\mu+1\right)^2+\left(n+\mu+\tfrac{1}{2}\right)^2-\gamma^2-\left(\mu+\tfrac{1}{2}\right)^2$$
$$\Sigma_{n,n+1}=\Sigma_{n+1,n}=-\left(n+\gamma+\mu+1\right)\sqrt{(n+1)(n+2\mu+2)} \tag{54}$$

Now, the matrix wave equation (5) could be rewritten as $E|P\rangle=\Omega^{-1}\mathcal{H}|P\rangle=\mathcal{H}|P\rangle$ since $\Omega_{n,m}=\delta_{n,m}$. Hence, we conclude that the Hamiltonian matrix is $\mathcal{H}=\frac{1}{2}\lambda^2\Sigma$. Finally, we obtain the matrix elements of the potential function $\tilde{V}(x)$ in the basis (50) as $\tilde{\mathcal{V}}=\mathcal{H}-\tilde{\mathcal{T}}$ where $\tilde{\mathcal{T}}$ is the tridiagonal part of $\mathcal{T}$ given by the three terms with square and curly brackets in Eq. (51). Figure 8 is a plot of the potential function $V(x)$ (in units of $\frac{1}{2}\lambda^2$) for a given set of values of the physical parameters $\{V_2,\gamma,\mu\}$. The second and third methods produce stable results for any value of the basis size *N*. Nonetheless, these two results are not identical. On the other hand, the first and fourth methods produce results that vary with the size of the basis but both agree to a certain extent with the second method for a properly chosen basis size *N*.

**5.2.2 *Wilson system*:**

The wavefunction for this system is written in the standard format (1) and in terms of the four-parameter Wilson polynomial as follows

$$\psi(E,x)=\sqrt{\rho^{\gamma}(\varepsilon;\kappa;a,b)}\sum_{n=0}^{\infty}W_n^{\gamma}(\varepsilon;\kappa;a,b)\phi_n(x)\,, \tag{55}$$

where $W_n^{\gamma}(\varepsilon;\kappa;a,b)$ is defined in Appendix B by Eq. (B18) and $\rho^{\gamma}(\varepsilon;\kappa;a,b)$ is its weight function (B19). Moreover, we choose the physical parameters in the Wilson polynomial as $a=b$, $\gamma=\kappa$ and $\varepsilon=2E/\lambda^2$. Therefore, the recursion relation (B20) becomes $\varepsilon|P\rangle=\Sigma|P\rangle$, where

$$\begin{aligned}\Sigma_{n,n}&=\frac{1}{2}\left[\left(n+\gamma+a-\tfrac{1}{2}\right)^2-\left(\gamma-\tfrac{1}{2}\right)^2-\left(a-\tfrac{1}{2}\right)^2+\tfrac{1}{4}\right]\\ \Sigma_{n,n+1}=\Sigma_{n+1,n}&=-\frac{1}{4}\left(n+\gamma+a\right)\sqrt{\frac{(n+1)(n+2\gamma)(n+2a)(n+2\gamma+2a-1)}{(n+\gamma+a)^2-\frac{1}{4}}}\end{aligned} \tag{56}$$

Similarly, compatibility of the matrix wave equation with the three-term recursion relation of the Wilson polynomial give the Hamiltonian matrix as $\mathcal{H}=\frac{1}{2}\lambda^2\Sigma$. Finally, we obtain the matrix elements of the potential function $\tilde{V}(x)$ in the basis (50) as $\tilde{\mathcal{V}}=\mathcal{H}-\tilde{\mathcal{T}}$ where $\tilde{\mathcal{T}}$ is the tridiagonal part of $\mathcal{T}$ given by the terms with square and curly brackets in Eq. (51). Figure 9 is a plot of the potential function $V(x)$ (in units of $\frac{1}{2}\lambda^2$) for a given set of values of the physical parameters $\{V_2,\gamma,\mu,a\}$. The second and third methods produce stable results for any value of the basis size *N*. Nonetheless, these two results are not identical. On the other hand, the first and fourth methods produce results that vary with the size of the basis and none agrees with the other two methods.

We end this subsection with a comment on the accuracy of the numerical results obtained by the four methods as shown in Fig. 8 and Fig. 9. Recently, we considered the solution of the wave equation in the "Tridiagonal Representation Approach" [9] where we encountered a potential function with the following properties:

- One of its terms is $\dfrac{V_2}{\sinh^2(\lambda x)}$, and
- The expansion coefficients of the corresponding wavefunction in the basis (50) satisfy a three-term recursion relation with recursion coefficients that are almost identical to the matrix elements of the kinetic energy operator in Eq. (51).

That potential is $V(x)=\dfrac{V_2}{\sinh^2(\lambda x)}+\dfrac{V_1}{\cosh^2(\lambda x)}+V_0$, which is the hyperbolic Pöschl-Teller potential. We found a perfect match between the result of the second method and this potential for a proper choice of potential parameters $V_1$ and $V_0$.

## 6. Conclusion

One of the advantages of reformulating quantum mechanics without a potential function is to enlarge the class of analytically describable systems (i.e., exactly solvable problems) even if no corresponding potential function could be realized. In this reformulation, orthogonal polynomials in the energy variable contain all the physical information about the system. Thus, the properties of these polynomials (e.g., weight function, asymptotics, recursion relation, distribution of zeros, orthogonality, etc.) determine the features of the physical system. To establish a correspondence between the new formulation and the conventional one, we tried to reconstruct the potential function using only elements of the new formulation like the recursion relation of the energy polynomials and the $L^2$ basis used. There is no a priori guarantee that such reconstruction could be achieved unless a stringent requirement is placed on the basis and the sought after potential function whereby the matrix representation of the wave operator in the chosen basis is tridiagonal and symmetric. If so, then we will be able to recover the potential function either exactly by analytic means or approximately by numerical schemes. In this work, we did establish such procedure that was proved viable.

## **Acknowledgements**:

The support by the Saudi Center for Theoretical Physics (SCTP) during the progress of this work is highly appreciated.

## Appendix A: Gauss Quadrature

Let $\{p_n(x)\}_{n=0}^{\infty}$ be a complete set of orthonormal polynomials over some interval in configuration space $x\in[x_-,x_+]$ with $\rho(x)$ being the normalized weight function. That is,

$$\int_{x_-}^{x_+}\rho(x)\,p_n(x)\,p_m(x)dx=\delta_{nm}\,. \tag{A1}$$

They satisfy the following symmetric three-term recursion relation

$$x\,p_n(x)=a_np_n(x)+b_{n-1}p_{n-1}(x)+b_np_{n+1}(x),\quad \text{for } n=1,2,3,..., \tag{A2}$$

with $b_n\neq 0$, $p_0(x)=1$ and $p_1(x)=\alpha x+\beta$. If $\alpha=b_0^{-1}$ and $\beta=-a_0b_0^{-1}$, then we call these the "polynomials of the first kind". Let us construct the $N\times N$ tridiagonal symmetric matrix

$$J=\begin{pmatrix} a_0 & b_0 & & & & & \\ b_0 & a_1 & b_1 & & & & \\ & b_1 & a_2 & b_2 & & & \\ & & b_2 & a_3 & \times & & \\ & & & \times & \times & \times & \\ & & & & \times & \times & \times \\ & & & & & \times & \times \end{pmatrix}, \tag{A3}$$

and let $\{\tau_n\}_{n=0}^{N-1}$ be its distinct eigenvalues with the corresponding normalized eigenvectors $\{\Lambda_{mn}\}_{m=0}^{N-1}$. Then, one can show that

$$p_m(\tau_n)=\Lambda_{mn}/\Lambda_{0n}\,,\quad n,m=0,1,...,N-1\,. \tag{A4}$$

Moreover, $\{\tau_n\}_{n=0}^{N-1}$ are the zeros of the polynomial of the first kind $p_N(x)$. Now, Gauss quadrature integral approximation states that if a function *f*(*x*) is integrable in the interval $x\in[x_-,x_+]$ with respect to the measure $\rho(x)dx$, then

$$\int_{x_-}^{x_+} \rho(x) f(x) dx \cong \sum_{n=0}^{N-1} \omega_n f(\tau_n), \qquad \text{(A5)}$$

where $\{\omega_n\}_{n=0}^{N-1}$ are referred to as the "numerical weights" associated with the quadrature, which could be computed either as $\omega_n = \Lambda_{0n}^2$ or in terms of the eigenvalues as

$$\omega_n = \frac{\prod_{m=0}^{N-2} \tau_n - \hat{\tau}_m}{\prod_{\substack{k=0 \\ k\neq n}}^{N-1} \tau_n - \tau_k}, \qquad \text{(A6)}$$

where $\{\hat{\tau}_m\}_{m=0}^{N-2}$ is the set of eigenvalues of the submatrix $\hat{J}$ obtained from $J$ by deleting the top (zeroth) row and left (zeroth) column.

Let the configuration space coordinate $r$ be related to the polynomial variable $x$ by $x = x(\lambda r)$, where $\lambda$ is a length scale parameter. Now, many of the basis elements in this configuration space are written in terms of orthonormal polynomials as $\phi_n(r) = \sqrt{\gamma(x)\rho(x)}\, p_n(x)$, where $\gamma(x)$ is some proper function. Then, using the integral measure $\lambda\int ...\, dr = \int ... \,(x')^{-1} dx$ where $x'(x) = \lambda^{-1}(dx/dr)$ and the orthogonality (A1), we find that $\bar{\phi}_n(x) = [x'(x)/\gamma(x)]\phi_n(x) = x'(x)\sqrt{\rho(x)/\gamma(x)}\, p_n(x)$. For example, in the 1D basis (10), $x' = 1$, $p_n(x) = [2^n n!]^{-1/2} H_n(x)$, $\rho(x) = \pi^{-1/2} e^{-x^2}$ and $\gamma(x) = 1$. Whereas, in the 3D basis (7), $x' = 1$, $p_n(x) = \sqrt{\frac{\Gamma(n+1)\Gamma(2\ell+2)}{\Gamma(n+2\ell+2)}}\, L_n^{2\ell+1}(x)$, $\rho(x) = x^{2\ell+1} e^{-x} / \Gamma(2\ell+2)$ and $\gamma(x) = x$. In such bases, the integral (15) in Sec. 3.1 could be evaluated as follows

$$\begin{aligned} \lambda\int \bar{\phi}_m(r) dr &= \int \rho(x) \frac{p_m(x)}{\sqrt{\gamma(x)\rho(x)}} dx \cong \sum_{n=0}^{N-1} \omega_n \frac{p_m(\tau_n)}{\sqrt{\gamma(\tau_n)\rho(\tau_n)}} \\ &= \sum_{n=0}^{N-1} \frac{\Lambda_{mn}\Lambda_{0n}}{\sqrt{\gamma(\tau_n)\rho(\tau_n)}} = \left(\Lambda W \Lambda^{\mathrm{T}}\right)_{m0} \end{aligned} \qquad \text{(A7)}$$

where $W$ is $N\times N$ diagonal matrix with elements $W_{nn} = 1\big/\sqrt{\gamma(\tau_n)\rho(\tau_n)}$. Consequently, Eq. (16) could be rewritten as

$$V(r) \cong \sum_{n,m=0}^{N-1} \bar{\phi}_n(r) \mathcal{V}_{nm} \left(\Lambda W \Lambda^{\mathrm{T}}\right)_{m0} = \sum_{n=0}^{N-1} U_n \bar{\phi}_n(r), \qquad \text{(A8)}$$

where $U_n = \left(\mathcal{V}\Lambda W \Lambda^{\mathrm{T}}\right)_{n0}$.

## Appendix B: Relevant energy polynomials

For ease of reference, we define in this Appendix the three orthogonal energy polynomials that are relevant to our current study and give their main properties. We choose the orthonormal version of these polynomials and give their discrete versions that are used as expansion coefficients of the bound states.

**B.1 The two-parameter Meixner-Pollaczek polynomial:**
The orthonormal version of this polynomial is written as follows (see pages 37-38 of Ref. [10])

$$P_n^{\mu}(y,\theta) = \sqrt{\frac{\Gamma(n+2\mu)}{\Gamma(2\mu)\Gamma(n+1)}}\; e^{in\theta}\,{}_2F_1\left(\begin{matrix}-n,\mu+iy\\2\mu\end{matrix}\middle|1-e^{-2i\theta}\right), \tag{B1}$$

where $y$ is the whole real line, $\mu > 0$ and $0 < \theta < \pi$. This is a polynomial in $y$ which is orthonormal with respect to the measure $\rho^{\mu}(y,\theta)dy$. That is,

$$\int_{-\infty}^{+\infty} P_n^{\mu}(y,\theta)P_m^{\mu}(y,\theta)\rho^{\mu}(y,\theta)dy = \delta_{nm}, \tag{B2}$$

where the normalized weight function is

$$\rho^{\mu}(y,\theta) = \frac{1}{2\pi\Gamma(2\mu)}(2\sin\theta)^{2\mu}e^{(2\theta-\pi)y}\left|\Gamma(\mu+iy)\right|^2. \tag{B3}$$

These polynomials satisfy the following symmetric three-term recursion relation

$$\begin{aligned}\left(y\sin\theta\right)P_n^{\mu}(y,\theta) &= -\left[\left(n+\mu\right)\cos\theta\right]P_n^{\mu}(y,\theta)\\ &+\tfrac{1}{2}\sqrt{n(n+2\mu-1)}P_{n-1}^{\mu}(y,\theta)+\tfrac{1}{2}\sqrt{(n+1)(n+2\mu)}P_{n+1}^{\mu}(y,\theta)\end{aligned} \tag{B4}$$

The generating function associated with these polynomials is written as

$$\sum_{n=0}^{\infty}\tilde{P}_n^{\mu}(y,\theta)t^n = \left(1-te^{i\theta}\right)^{-\mu+iy}\left(1-te^{-i\theta}\right)^{-\mu-iy}. \tag{B5}$$

where $\tilde{P}_n^{\mu}(y;\theta) = \sqrt{\frac{\Gamma(n+2\mu)}{\Gamma(2\mu)\Gamma(n+1)}}P_n^{\mu}(y;\theta)$. The asymptotics ($n\to\infty$) is (see, for example, the Appendix in Ref. [2])

$$P_n^{\mu}(y;\theta) \approx \frac{2n^{-1/2}e^{(\pi/2-\theta)y}}{(2\sin\theta)^{\mu}\left|\Gamma(\mu+iy)\right|}\cos\left[n\theta+\arg\Gamma(\mu+iy)-\mu\tfrac{\pi}{2}-y\ln(2n\sin\theta)\right], \tag{B6}$$

which is in the required general form given by Eq. (3) because the $n$-dependent term $y\ln(2n\sin\theta)$ in the argument of the cosine could be ignored relative to $n\theta$ since $\ln n \approx o(n)$ as $n\to\infty$. The scattering amplitude in the asymptotics (B6) shows that a discrete infinite spectrum occur if $\mu+iy=-m$, where $m=0,1,2,\ldots$ Thus, the spectrum formula associated with this polynomial is $y^2=-(m+\mu)^2$ and bound states will be written as in Eq. (4) where the discrete version of this polynomial is the Meixner polynomial whose normalized version is written as (see pages 45-46 in Ref. [10])

$$M_n^{\mu}\left(m;\beta\right) = \sqrt{\frac{\Gamma(n+2\mu)\,\beta^n}{\Gamma(2\mu)\Gamma(n+1)}}\,{}_2F_1\left(\begin{matrix}-n,-m\\2\mu\end{matrix}\middle|1-\beta^{-1}\right), \tag{B7}$$

where $0<\beta<1$. It satisfies the following recursion relation

$$\begin{aligned}(\beta-1)m\,M_n^{\mu}\left(m;\beta\right) &= -\left[n(1+\beta)+2\mu\beta\right]M_n^{\mu}\left(m;\beta\right)\\ &+\sqrt{n(n+2\mu-1)\beta}\,M_{n-1}^{\mu}\left(m;\beta\right)+\sqrt{(n+1)(n+2\mu)\beta}\,M_{n+1}^{\mu}\left(m;\beta\right)\end{aligned} \tag{B8}$$

The associated normalized discrete weight function is $\rho_m^{\mu}(\beta) = (1-\beta)^{2\mu}\frac{\Gamma(m+2\mu)\,\beta^m}{\Gamma(2\mu)\Gamma(m+1)}$. That is, $\sum_{m=0}^{\infty}\rho_m^{\mu}(\beta)\,M_n^{\mu}\left(m;\beta\right)M_{n'}^{\mu}\left(m;\beta\right) = \delta_{n,n'}$.

**B.2 The three-parameter continuous dual Hahn polynomial:**
The normalized version of this polynomial is (see pages 29-31 of Ref. [10])

$$S_n^{\mu}(y^2;a,b) = \sqrt{\frac{(\mu+a)_n(\mu+b)_n}{n!\,(a+b)_n}}\;{}_3F_2\left(\begin{matrix}-n,\mu+iy,\mu-iy\\\mu+a,\mu+b\end{matrix}\middle|1\right), \tag{B9}$$

where $(z)_n = z(z+1)(z+2)...(z+n-1) = \frac{\Gamma(n+z)}{\Gamma(z)}$ and ${}_3F_2\left(\begin{matrix} a,b,c \\ d,e \end{matrix}\middle| z\right) = \sum_{n=0}^{\infty} \frac{(a)_n(b)_n(c)_n}{(d)_n(e)_n} \frac{z^n}{n!}$ is the generalized hypergeometric function. $y > 0$ and the parameters $\{\mu, a, b\}$ are positive except for a pair of complex conjugates with positive real parts.. This is a polynomial in $y^2$ which is orthonormal with respect to the measure $\rho^\mu(y;a,b)dy$ where the weight function reads as follows

$$\rho^\mu(y;a,b) = \frac{1}{2\pi} \frac{\left|\Gamma(\mu+\mathrm{i}y)\Gamma(a+\mathrm{i}y)\Gamma(b+\mathrm{i}y)/\Gamma(2\mathrm{i}y)\right|^2}{\Gamma(\mu+a)\Gamma(\mu+b)\Gamma(a+b)}. \tag{B10}$$

That is, $\int_0^\infty S_n^\mu(y^2;a,b)S_m^\mu(y^2;a,b)\rho^\mu(y;a,b)dy = \delta_{nm}$. It also satisfies the following symmetric three-term recursion relation

$$\begin{aligned} y^2 S_n^\mu &= \left[(n+\mu+a)(n+\mu+b) + n(n+a+b-1) - \mu^2\right] S_n^\mu \\ &\quad -\sqrt{n(n+a+b-1)(n+\mu+a-1)(n+\mu+b-1)}\, S_{n-1}^\mu \\ &\quad -\sqrt{(n+1)(n+a+b)(n+\mu+a)(n+\mu+b)}\, S_{n+1}^\mu \end{aligned} \tag{B11}$$

The generating function is written as follows

$$\sum_{n=0}^{\infty} \tilde{S}_n^\mu(z^2;a,b)t^n = (1-t)^{-\mu+\mathrm{i}z}\, {}_2F_1\left(\begin{matrix} a+\mathrm{i}z, b+\mathrm{i}z \\ a+b \end{matrix}\middle| t\right), \tag{B12}$$

where $\tilde{S}_n^\mu(z^2;a,b) = \frac{(\mu+a)_n(\mu+b)_n}{n!(a+b)_n}\, {}_3F_2\left(\begin{matrix} -n, \mu+\mathrm{i}z, \mu-\mathrm{i}z \\ \mu+a, \mu+b \end{matrix}\middle| 1\right)$. The asymptotics ($n \to \infty$) is (see, for example, the Appendix in Ref. [2])

$$\begin{aligned} S_n^\mu(y^2;a,b) &\approx \frac{2\sqrt{\Gamma(\mu+a)\Gamma(\mu+b)\Gamma(a+b)}\,\left|\Gamma(2\mathrm{i}y)\right|}{\left|\Gamma(\mu+\mathrm{i}y)\Gamma(a+\mathrm{i}y)\Gamma(b+\mathrm{i}y)\right|\sqrt{n}} \times \\ &\cos\left[y\ln n + \arg\Gamma(\mu+\mathrm{i}y) + \arg\Gamma(a+\mathrm{i}y) + \arg\Gamma(b+\mathrm{i}y) - \arg\Gamma(2\mathrm{i}y)\right] \end{aligned} \tag{B13}$$

Noting that $\ln n \approx o(n^\xi)$ for any $\xi > 0$, then this result is also in the required general form given by Eq. (3). The scattering amplitude in this asymptotics shows that a discrete finite spectrum occur if $\mu + \mathrm{i}y = -m$, where $m = 0,1,2,..,N$ and $N$ is the largest integer less than or equal to $-\mu$. Thus, the spectrum formula associated with this polynomial is $y^2 = -(m+\mu)^2$ and the discrete version of this polynomial is the dual Hahn polynomial whose normalized version is written as (see pages 34-36 in Ref. [10])

$$R_n^N(m;a,b) = \sqrt{\frac{(a+1)_n(b+1)_{N-n}}{n!(N-n)!}}\, {}_3F_2\left(\begin{matrix} -n, -m, m+a+b+1 \\ a+1, -N \end{matrix}\middle| 1\right), \tag{B14}$$

where $n, m = 0,1,2,..,N$ and either $a, b > -1$ or $a, b < -N$. It satisfies the following recursion relation

$$\begin{aligned} \left(m + \tfrac{a+b+1}{2}\right)^2 R_n^N &= -\left\{\left(n + \tfrac{a+1}{2}\right)^2 + \left(n - \tfrac{b+1}{2}\right)^2 - N(2n+a+1) \right. \\ &\quad \left. -\tfrac{1}{4}\left[(a+b+1)^2 + (a+1)^2 + (b+1)^2\right]\right\} R_n^N \\ &+\sqrt{n(n+a)(N-n+1)(N-n+b+1)}\,R_{n-1}^N + \sqrt{(n+1)(n+a+1)(N-n)(N-n+b)}\,R_{n+1}^N \end{aligned} \tag{B15}$$

The associated normalized discrete weight function is

$$\rho^N(m;a,b) = (N!)\frac{(2m+a+b+1)(a+1)_m(N-m+1)_m}{(m+a+b+1)_{N+1}(b+1)_m m!}. \tag{B16}$$

Therefore, we can write $\sum_{m=0}^{N}\rho^{N}(m;a,b)\ R_n^N(m;a,b)\,R_{n'}^N(m;a,b)=\delta_{n,n'}$.

**B.3 The four-parameter Wilson polynomial:**

The Wilson polynomial, $\tilde{W}_n^{\mu}(y^2;\nu;a,b)$, is defined here as (see pages 24-26 in Re. [10])

$$\tilde{W}_n^{\mu}(y^2;\nu;a,b)=\frac{(\mu+a)_n(\mu+b)_n}{(a+b)_n n!}\,{}_4F_3\left(\begin{matrix}-n,n+\mu+\nu+a+b-1,\mu+iy,\mu-iy\\ \mu+\nu,\mu+a,\mu+b\end{matrix}\middle|1\right), \tag{B17}$$

where $y>0$ and the parameters $\{\mu,\nu,a,b\}$ are positive except for a pair of complex conjugates with positive real parts. This is a polynomial in $y^2$ whose orthonormal version is written as

$$W_n^{\mu}(y^2;\nu;a,b)=\sqrt{\left(\frac{2n+\mu+\nu+a+b-1}{n+\mu+\nu+a+b-1}\right)\frac{(\mu+\nu)_n(a+b)_n(\mu+\nu+a+b)_n n!}{(\mu+a)_n(\mu+b)_n(\nu+a)_n(\nu+b)_n}}\ \tilde{W}_n^{\mu}(y^2;\nu;a,b)\,. \tag{B18}$$

It is orthonormal with respect to the measure $\rho^{\mu}(y;a,b)dy$ where the normalized weight function reads as follows

$$\rho^{\mu}(y;\nu;a,b)=\frac{1}{2\pi}\frac{\Gamma(\mu+\nu+a+b)\left|\Gamma(\mu+iy)\Gamma(\nu+iy)\Gamma(a+iy)\Gamma(b+iy)/\Gamma(2iy)\right|^2}{\Gamma(\mu+\nu)\Gamma(a+b)\Gamma(\mu+a)\Gamma(\mu+b)\Gamma(\nu+a)\Gamma(\nu+b)}\,. \tag{B19}$$

It also satisfies the following symmetric three-term recursion relation

$$\begin{aligned}y^2W_n^{\mu}=&\left[\frac{(n+\mu+\nu)(n+\mu+a)(n+\mu+b)(n+\mu+\nu+a+b-1)}{(2n+\mu+\nu+a+b)(2n+\mu+\nu+a+b-1)}+\frac{n(n+\nu+a-1)(n+\nu+b-1)(n+a+b-1)}{(2n+\mu+\nu+a+b-1)(2n+\mu+\nu+a+b-2)}-\mu^2\right]W_n^{\mu}\\ &-\frac{1}{2n+\mu+\nu+a+b-2}\sqrt{\frac{n(n+\mu+\nu-1)(n+a+b-1)(n+\mu+a-1)(n+\mu+b-1)(n+\nu+a-1)(n+\nu+b-1)(n+\mu+\nu+a+b-2)}{(2n+\mu+\nu+a+b-3)(2n+\mu+\nu+a+b-1)}}\,W_{n-1}^{\mu}\\ &-\frac{1}{2n+\mu+\nu+a+b}\sqrt{\frac{(n+1)(n+\mu+\nu)(n+a+b)(n+\mu+a)(n+\mu+b)(n+\nu+a)(n+\nu+b)(n+\mu+\nu+a+b-1)}{(2n+\mu+\nu+a+b-1)(2n+\mu+\nu+a+b+1)}}\,W_{n+1}^{\mu}\end{aligned} \tag{B22}$$

The generating function is written as follows

$$\sum_{n=0}^{\infty}\tilde{W}_n^{\mu}(y^2;\nu;a,b)t^n={}_2F_1\left(\begin{matrix}\mu+iy,\nu+iy\\ \mu+\nu\end{matrix}\middle|t\right){}_2F_1\left(\begin{matrix}a-iy,b-iy\\ a+b\end{matrix}\middle|t\right), \tag{B21}$$

The asymptotics ($n\to\infty$) is (see, for example, Appendix B in Ref. [3])

$$W_n^{\mu}(y^2;\nu;a,b)\approx 2\sqrt{\frac{2}{n}}B(\mu,\nu,a,b)\left|\mathcal{A}(iy)\right|\cos\left[2y\ln n+\arg\mathcal{A}(iy)\right], \tag{B22}$$

where $B(\mu,\nu,a,b)=\sqrt{\Gamma(\mu+\nu)\Gamma(a+b)\Gamma(\mu+a)\Gamma(\mu+b)\Gamma(\nu+a)\Gamma(\nu+b)/\Gamma(\mu+\nu+a+b)}$ and $\mathcal{A}(z)=\Gamma(2z)/\Gamma(\mu+z)\Gamma(\nu+z)\Gamma(a+z)\Gamma(b+z)$. Again, noting that $\ln n\approx o(n^{\xi})$ for any $\xi>0$, then this result is also in the required general form given by Eq. (3). The scattering amplitude in this asymptotics shows that a discrete finite spectrum occur if $\mu+\mathrm{i}y=-m$, where $m=0,1,2,..,N$ and $N$ is the largest integer less than or equal to $-\mu$. Thus, the spectrum formula associated with this polynomial is $y^2=-(m+\mu)^2$ and the discrete version of this polynomial is the Racah polynomial which we can write as (see pages 26-29 in Ref. [10])

$$\tilde{R}_n^N(m;\alpha,\beta,\gamma)=\frac{(\alpha+1)_n(\gamma+1)_n}{(\alpha+\beta+N+2)_n n!}\,{}_4F_3\left(\begin{matrix}-n,-m,n+\alpha+\beta+1,m-\beta+\gamma-N\\ \alpha+1,\gamma+1,-N\end{matrix}\middle|1\right) \tag{B23}$$

where $n,m=0,1,2,..,N$ and the parameter constraints are $\alpha>-1$, $\gamma>-1$, $\beta>N-1$. It satisfies the following recursion relation

$$\tfrac{1}{4}\left(N+\beta-\gamma-2m\right)^2\tilde{R}_n^N=$$
$$\left[\tfrac{1}{4}\left(N+\beta-\gamma\right)^2-\tfrac{(n-N)(n+\alpha+1)(n+\gamma+1)(n+\alpha+\beta+1)}{(2n+\alpha+\beta+1)(2n+\alpha+\beta+2)}-\tfrac{n(n+\beta)(n+\alpha+\beta-\gamma)(n+N+\alpha+\beta+1)}{(2n+\alpha+\beta)(2n+\alpha+\beta+1)}\right]\tilde{R}_n^N \quad \text{(B24)}$$
$$+\tfrac{(n+\alpha)(n+\beta)(n+\gamma)(n+\alpha+\beta-\gamma)}{(2n+\alpha+\beta)(2n+\alpha+\beta+1)}\tilde{R}_{n-1}^N+\tfrac{(n+1)(n-N)(n+\alpha+\beta+1)(n+N+\alpha+\beta+2)}{(2n+\alpha+\beta+1)(2n+\alpha+\beta+2)}\tilde{R}_{n+1}^N$$

The discrete orthogonality relation reads as follows

$$\sum_{m=0}^{N}\frac{2m+\gamma-\beta-N}{m+\gamma-\beta-N}\frac{(-N)_m(\alpha+1)_m(\gamma+1)_m(\gamma-\beta-N+1)_m}{(-\beta-N)_m(\gamma-\beta+1)_m(\gamma-\alpha-\beta-N)_m m!}\bar{R}_n^N(m;\alpha,\beta,\gamma)\bar{R}_{n'}^N(m;\alpha,\beta,\gamma)$$
$$=\frac{n+\alpha+\beta+1}{2n+\alpha+\beta+1}\frac{(-\alpha-\beta-N-1)_N(\gamma-\beta-N+1)_N}{(-\beta-N)_N(\gamma-\alpha-\beta-N)_N}\frac{(\beta+1)_n(\alpha+\beta-\gamma+1)_n(\alpha+\beta+N+2)_n n!}{(-N)_n(\alpha+1)_n(\gamma+1)_n(\alpha+\beta+2)_n}\delta_{n,n'} \quad \text{(B25)}$$

where $\bar{R}_n^N(m;\alpha,\beta,\gamma)={}_4F_3\left(\begin{smallmatrix}-n,-m,n+\alpha+\beta+1,m-\beta+\gamma-N\\ \alpha+1,\gamma+1,-N\end{smallmatrix}\middle|1\right)$. Thus, the discrete normalized weight function is

$$\rho^N(m;\alpha,\beta,\gamma)=\frac{(-\beta-N)_N(\gamma-\alpha-\beta-N)_N}{(-\alpha-\beta-N-1)_N(\gamma-\beta-N+1)_N}\times$$
$$\frac{2m+\gamma-\beta-N}{m+\gamma-\beta-N}\frac{(-N)_m(\alpha+1)_m(\gamma+1)_m(\gamma-\beta-N+1)_m}{(-\beta-N)_m(\gamma-\beta+1)_m(\gamma-\alpha-\beta-N)_m m!} \quad \text{(B26)}$$

and the orthonormal version of the discrete Racah polynomial is

$$R_n^N(m;\alpha,\beta,\gamma)=\sqrt{\frac{2n+\alpha+\beta+1}{n+\alpha+\beta+1}\frac{(-N)_n(\alpha+1)_n(\gamma+1)_n(\alpha+\beta+2)_n}{(\beta+1)_n(\alpha+\beta-\gamma+1)_n(\alpha+\beta+N+2)_n n!}}\times$$
$${}_4F_3\left(\begin{smallmatrix}-n,-m,n+\alpha+\beta+1,m-\beta+\gamma-N\\ \alpha+1,\gamma+1,-N\end{smallmatrix}\middle|1\right) \quad \text{(B27)}$$

Thus, we can rewrite (B25) as

$$\sum_{m=0}^{N}\rho^N(m;\alpha,\beta,\gamma)R_n^N(m;\alpha,\beta,\gamma)R_{n'}^N(m;\alpha,\beta,\gamma)=\delta_{n,n'}\,. \quad \text{(B28)}$$

**B.4 New orthogonal polynomial:**

The orthogonal polynomial whose three-term recursion relation is given by either Eq. (44) or Eq. (49) is not found in the mathematics literature. Its properties (weight function, generating function, orthogonality, asymptotics, etc.) are yet to be derived analytically. Nonetheless, a generalized version of this polynomial, which is very relevant to various problems in physics, has already been encountered in the physics literature (see, for example, article [11] and references therein). This generalized polynomial is a four-parameter polynomial that was referred to as the "*dipole polynomial*" and is designated by $H_n^{(\mu,\nu)}(y;\alpha,\theta)$. It satisfies the following symmetric three-term recursion relation

$$\left(\cos\theta\right)H_n^{(\mu,\nu)}(y;\alpha,\theta)=D_{n-1}H_{n-1}^{(\mu,\nu)}(y;\alpha,\theta)+D_nH_{n+1}^{(\mu,\nu)}(y;\alpha,\theta)$$
$$\left\{y\sin\theta\left[\left(n+\tfrac{\mu+\nu+1}{2}\right)^2+\alpha\right]+C_n\right\}H_n^{(\mu,\nu)}(y;\alpha,\theta) \quad \text{(B29)}$$

where $0\le\theta\le\pi$, $C_n$ and $D_n$ are defined below Eq. (51) above. It is a polynomial of degree *n* in $y$ and in $\alpha$. The special cases of this polynomial in section 5.1 correspond to $\theta=\pi/2$, $\mu=\nu$ and $y=u_1^{-1}$. Moreover, in subsection 5.1.1 $\alpha=u_0-\varepsilon$ whereas in subsection 5.1.2 $\alpha=u_0-\tfrac{1}{4}$.

## Figures Caption

**Fig. 1**: The solid thick curve is the potential function as recovered by the four methods using a basis size of 20. The thin curve with empty circles is the exact potential function, $V(r)=5r^2e^{-r}$. The basis used is that of Eq. (7) with $\ell=1$, $\lambda=7$ and the potential matrix elements are calculated using the Gauss quadrature associated with the Laguerre polynomials.

**Fig. 2**: The dashed trace is the potential functions $V(r)=5r^2e^{-r}$ whereas the solid trace is its nonanalytic version $\bar{V}(r)$ defined by Eq. (27).

**Fig. 3**: The left (right) plots are the results of recovering the nonanalytic potential $\bar{V}(r)$ using the third (fourth) method. The size of the basis are 10 (top), 20 (middle) and 32 (bottom).

**Fig. 4**: The red solid trace is the result of calculating the Coulomb potential using the third method as compared to the exact potential in dotted blue trace. We took $Z=2$, $\ell=1$ and $\lambda=3$. The three parts of the figure correspond to basis size: 10, 20, 50.

**Fig. 5**: The red solid trace is the result of calculating the Morse potential using the first and fourth methods as compared to the exact potential in dotted blue trace. We took $\beta=5$, $\lambda=1$, $V_0=\lambda^2/8$ and $\alpha=3$. The basis size was taken 100 for the first method and 10 for the fourth method.

**Fig. 6**: Plot of the potential (42) obtained by varying the parameter $V_1$ while keeping the other two fixed at $V_0=0$ and $V_2=\lambda^2$. We took $\lambda=1$ and varied $V_1$ from 0 (top curve) to 5 (bottom curve) in units of $\lambda^2$. The *x*-axis is measured in units of $\pi/2\lambda$.

**Fig. 7**: Plot of the potential (47) obtained by varying the parameter $V_0$ while keeping the $V_1$ fixed at $V_1=1.5\lambda^2$. We took $\lambda=1$ and varied $V_0$ from −1 (bottom curve) to +1 (top curve) in units of $\lambda^2$.

**Fig. 8**: Plot of the potential function $V(x)$ of subsection 5.2.1 for the physical parameters $\{V_2,\gamma,\mu\}=\{\lambda^2,-10,3\}$. The second and third methods produce stable results for any value of the basis size *N*. However, these two results are not identical. The first and fourth methods produce results that vary with the size of the basis but both agree to a certain extent with the second method for a chosen basis size of 18.

**Fig. 9**: Plot of the potential function $V(x)$ of subsection 5.2.2 for the physical parameters $\{V_2,\gamma,\mu,a\}=\{\lambda^2,-7,2,2\}$. The second and third methods produce stable results for any value of the basis size *N*. However, these two results are not identical. The first and fourth methods produce results (not shown) that vary with the size of the basis and none agrees with the other two methods.

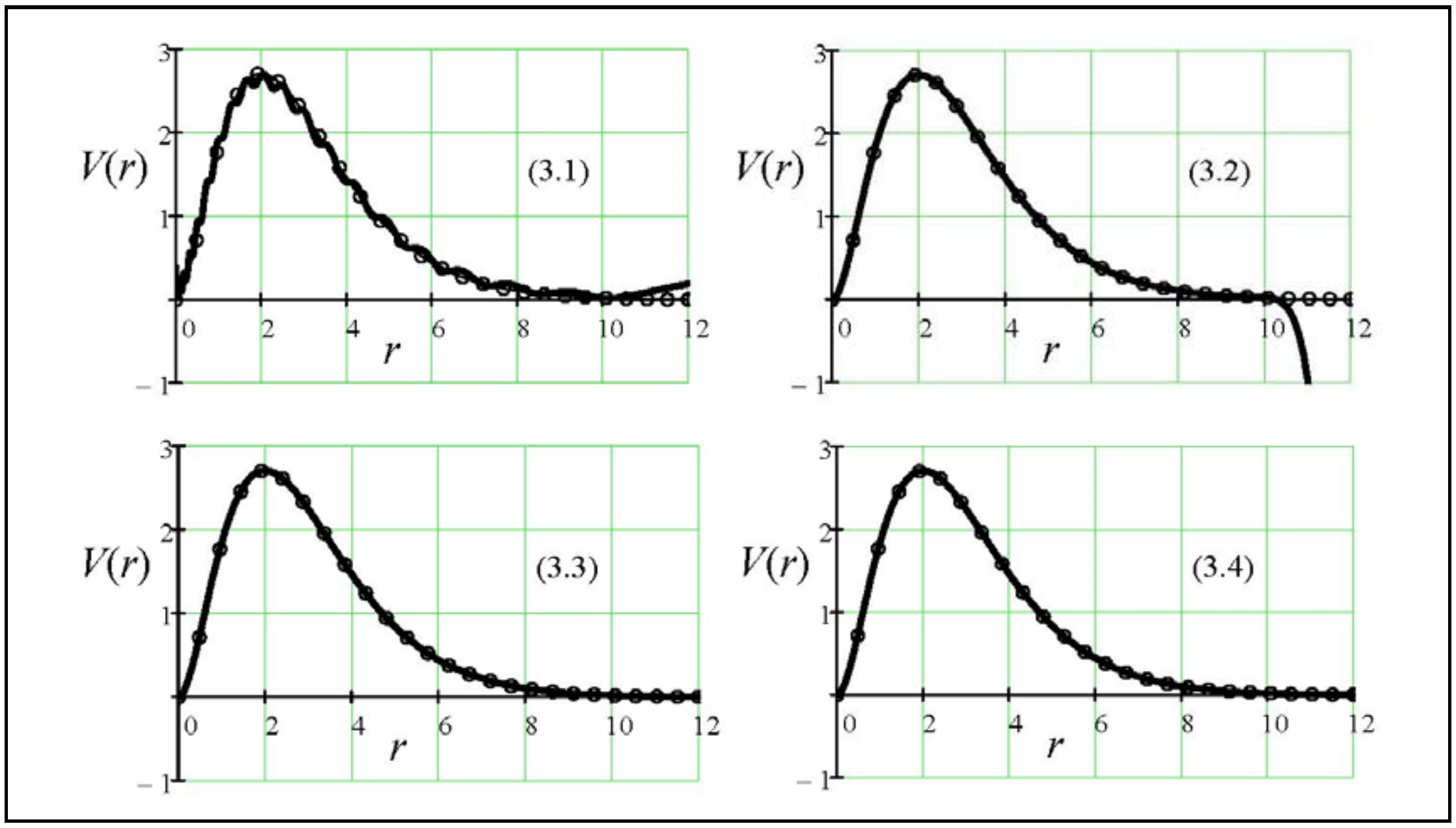

**Fig. 1**

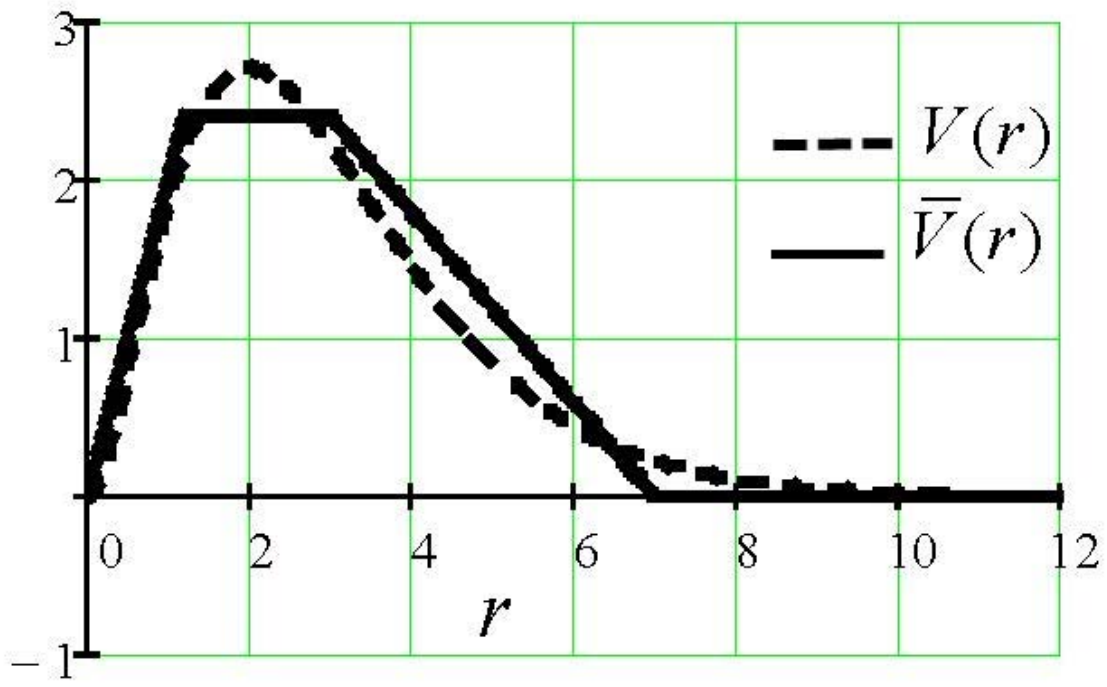

**Fig. 2**

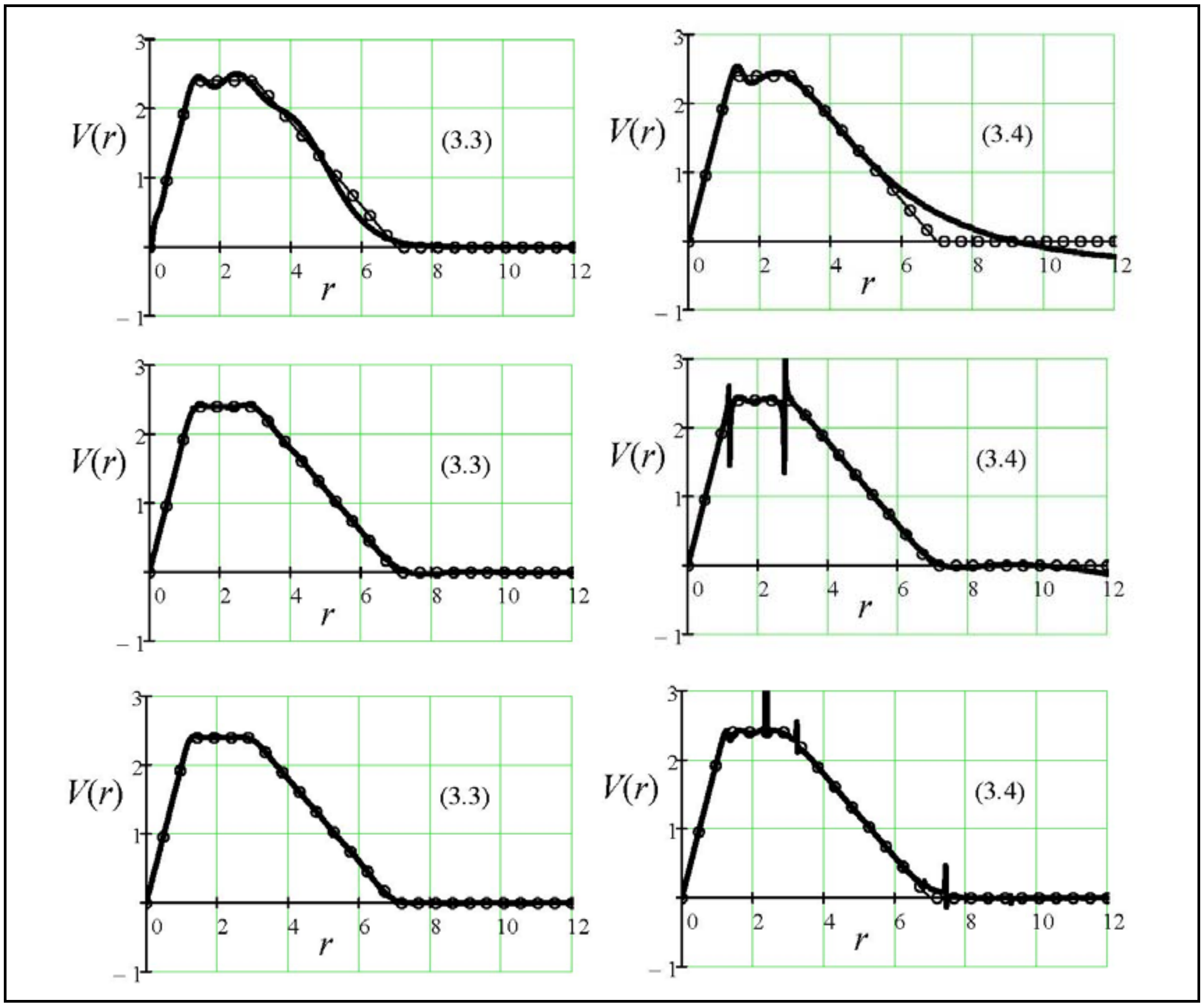

**Fig. 3**

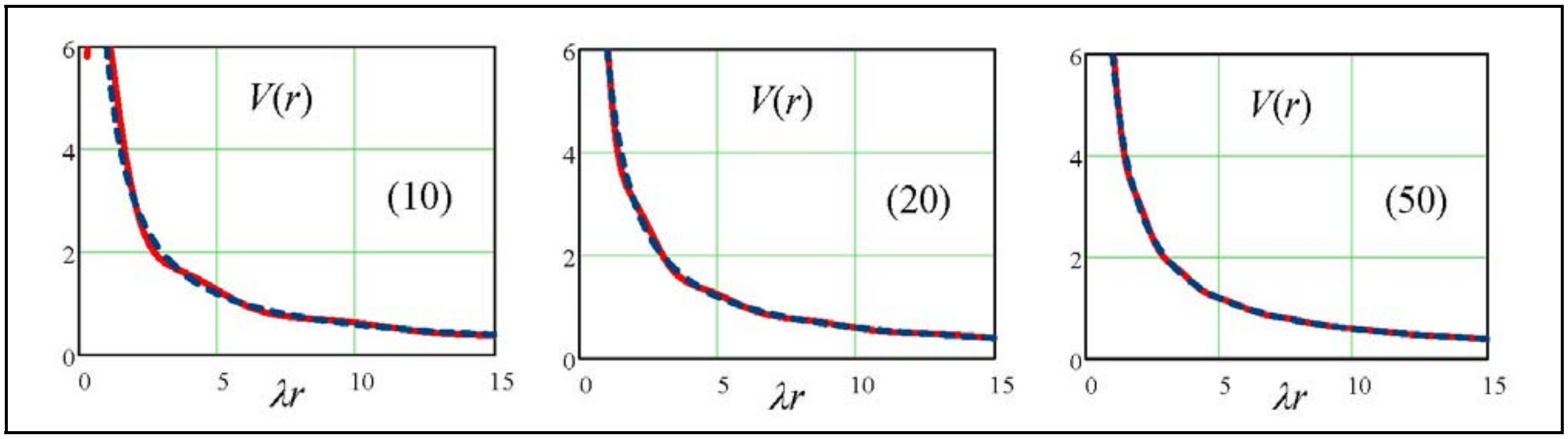

**Fig. 4**

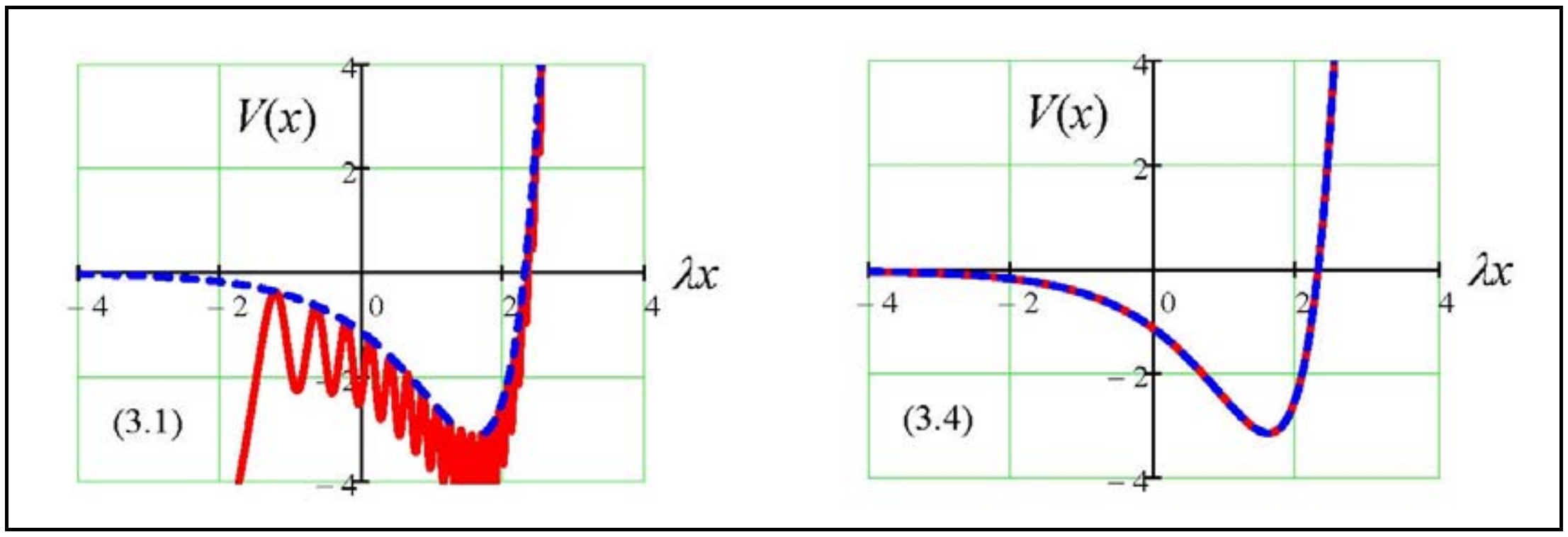

**Fig. 5**

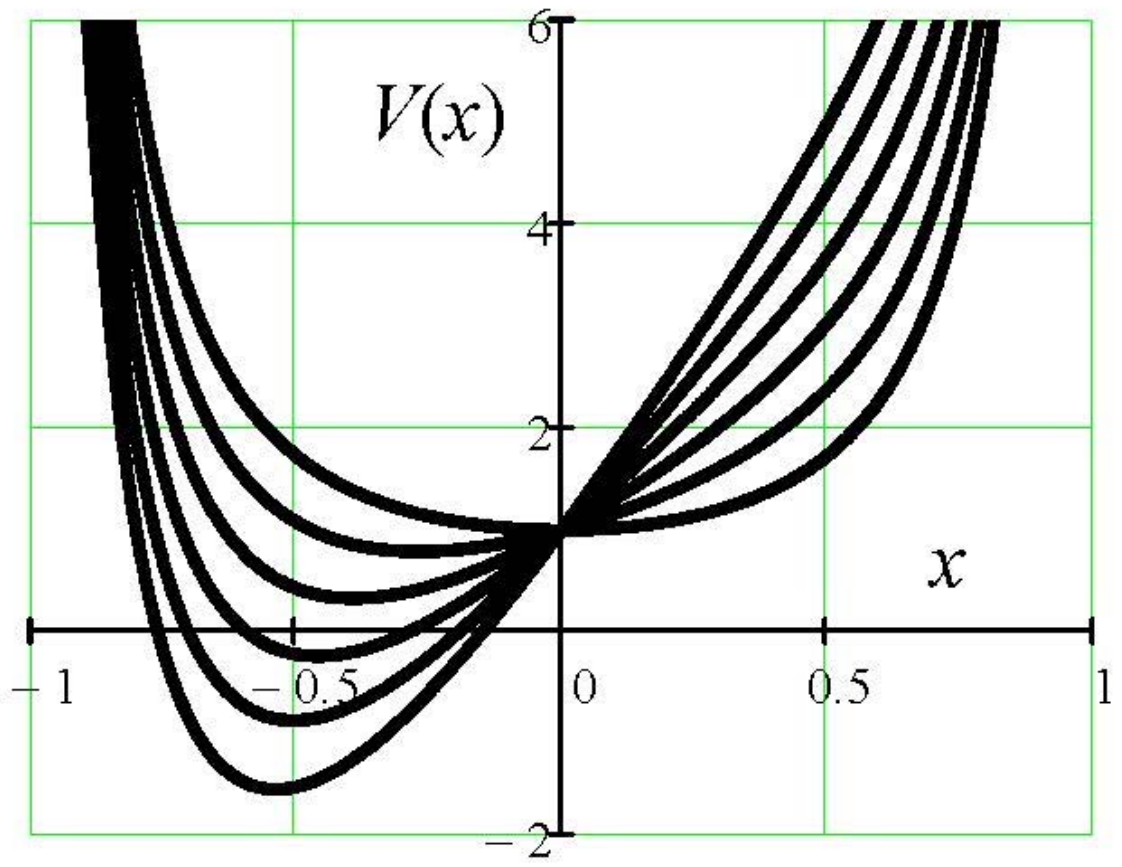

**Fig. 6**

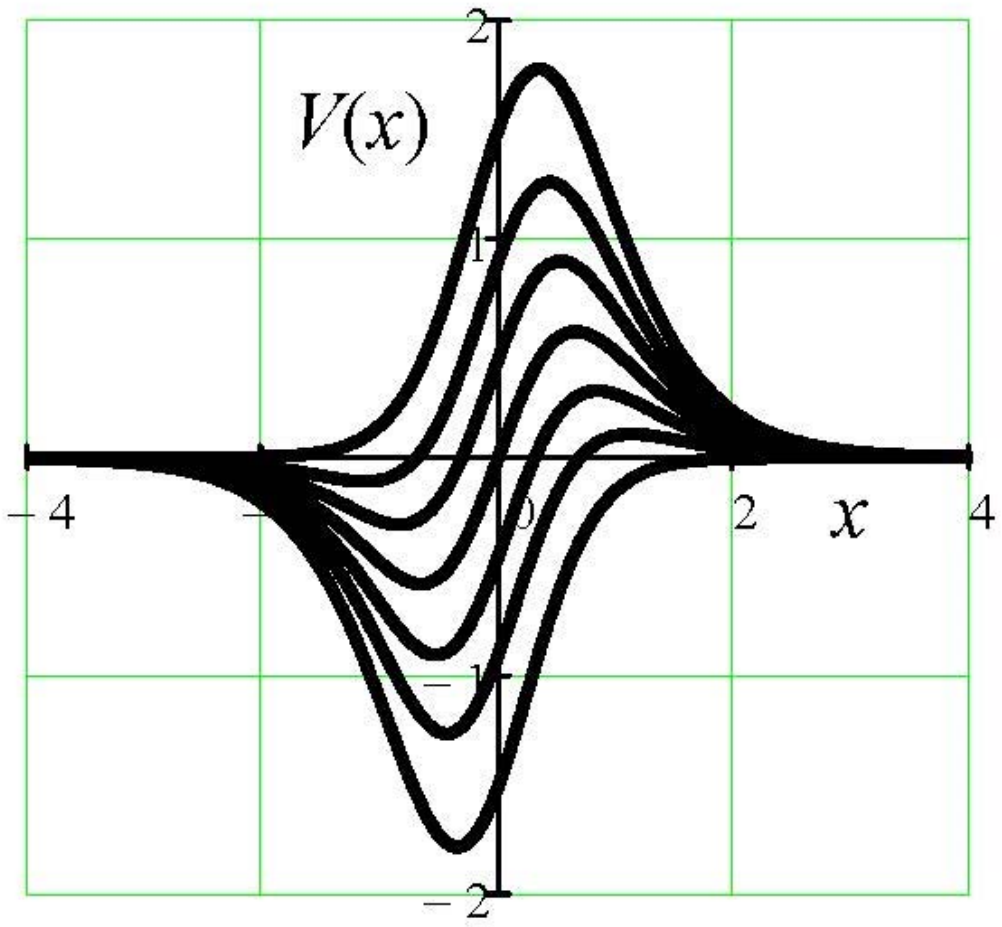

**Fig. 7**

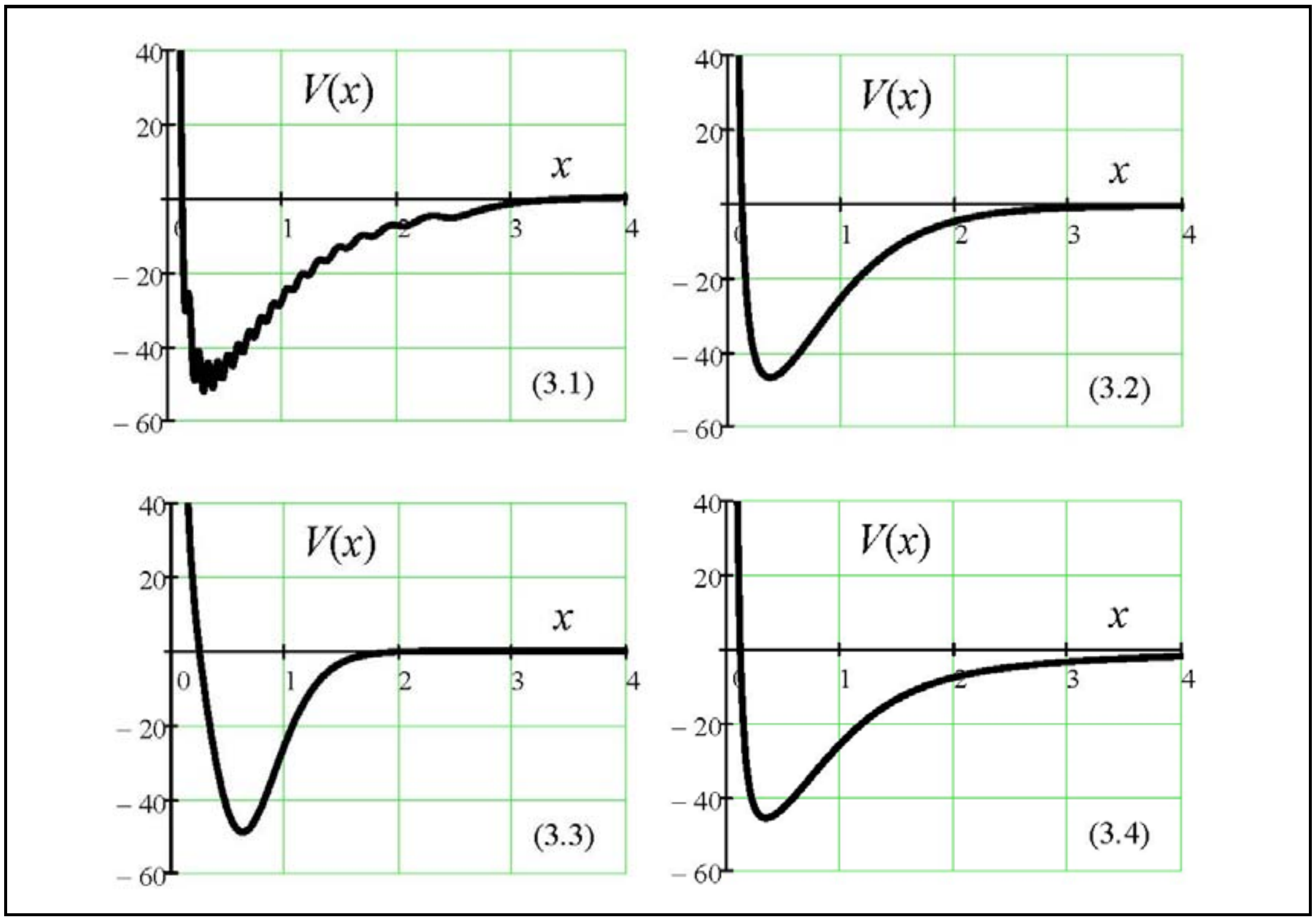

**Fig. 8**

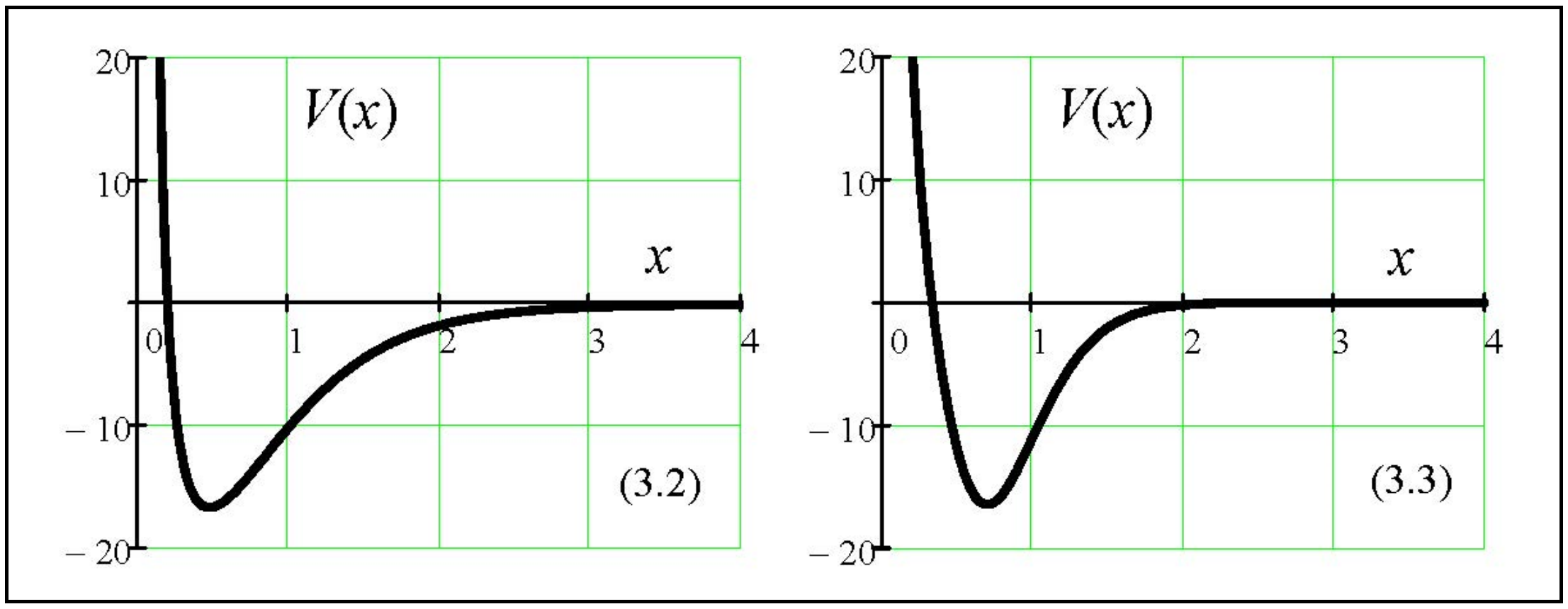

**Fig. 9**